\documentclass[journal]{IEEEtran}
\usepackage{amsmath,amsfonts}
\usepackage[noend]{algorithmic}
\usepackage{algorithm}
\usepackage{array}
\usepackage[caption=false,font=normalsize,labelfont=sf,textfont=sf]{subfig}
\usepackage{textcomp}
\usepackage{stfloats}
\usepackage{bm}
\usepackage{url}
\usepackage{verbatim}
\usepackage{graphicx}
\usepackage{cite}
\usepackage{marvosym}
\usepackage{booktabs}
\usepackage{tabularx}
\usepackage{multirow}
\usepackage{xcolor}
\begin{document}

\title{HashVFL: Defending Against Data Reconstruction Attacks in Vertical Federated Learning}

\author{Pengyu Qiu, Xuhong Zhang, Shouling Ji, Chong Fu, Xing Yang\textsuperscript{\Letter}, Ting Wang%
\IEEEcompsocitemizethanks{
\IEEEcompsocthanksitem P. Qiu, S. Ji, C. Fu are with the College of Computer Science and Technology at Zhejiang University, Hangzhou, Zhejiang, 310027, China. E-mail:\{qiupys,sji,fuchong\}@zju.edu.cn\protect\\
\IEEEcompsocthanksitem X. Zhang is with the School of Software Technology at Zhejiang University, Ningbo, Zhejiang, 315048, China. E-mail: zhangxuhong@zju.edu.cn\protect\\
\IEEEcompsocthanksitem X. Yang is with the Hefei Interdisciplinary Center, National University of Defense Technology, Hefei, Anhui, 230037, China. E-mail: yangxing17@nudt.edu.cn. He is also the corresponding author of this paper.\protect\\
\IEEEcompsocthanksitem T. Wang are with the College of Information Science and Technology at Pennsylvania State University, State College, PA, 16801, United States. E-mail: inbox.ting@gmail.com
}%
}

\maketitle

\begin{abstract}
Vertical Federated Learning (VFL) is a trending collaborative machine learning model training solution. 
Existing industrial frameworks employ secure multi-party computation techniques such as homomorphic encryption to ensure data security and privacy. 
Despite these efforts, studies have revealed that data leakage remains a risk in VFL due to the correlations between intermediate representations and raw data.
Neural networks can accurately capture these correlations, allowing an adversary to reconstruct the data. 
This emphasizes the need for continued research into securing VFL systems.

Our work shows that hashing is a promising solution to counter data reconstruction attacks. 
The one-way nature of hashing makes it difficult for an adversary to recover data from hash codes. 
However, implementing hashing in VFL presents new challenges, including vanishing gradients and information loss. 
To address these issues, we propose HashVFL, which integrates hashing and simultaneously achieves learnability, bit balance, and consistency.

Experimental results indicate that HashVFL effectively maintains task performance while defending against data reconstruction attacks. 
It also brings additional benefits in reducing the degree of label leakage, mitigating adversarial attacks, and detecting abnormal inputs.
We hope our work will inspire further research into the potential applications of HashVFL.
\end{abstract}


\section{Introduction}
\label{sec:introduction}
Machine learning algorithms, particularly Deep Neural Networks (DNNs), have seen significant growth in recent decades \cite{Breiman2001machine,Chang2011svm,LeCun2015deep}. 
DNNs have been applied in finance \cite{Dautel2020forex,Dixon2017markets}, biomedicine \cite{Jumper2021alpha,Wang2019clinical}, and even military operations \cite{Krebs2017object,Kupryashkin2022military}. 
However, data security and privacy are of the utmost importance in these sensitive fields, and strict laws and regulations like GDPR \cite{voigt2017gdpr} and CCPA \cite{kiselbach2012ccpa} limit the flow of data. 
This creates a dilemma between the need for large amounts of data in machine learning models and the restrictions on data flow.

\begin{figure}[t]
    \centering
    \includegraphics[width=\columnwidth]{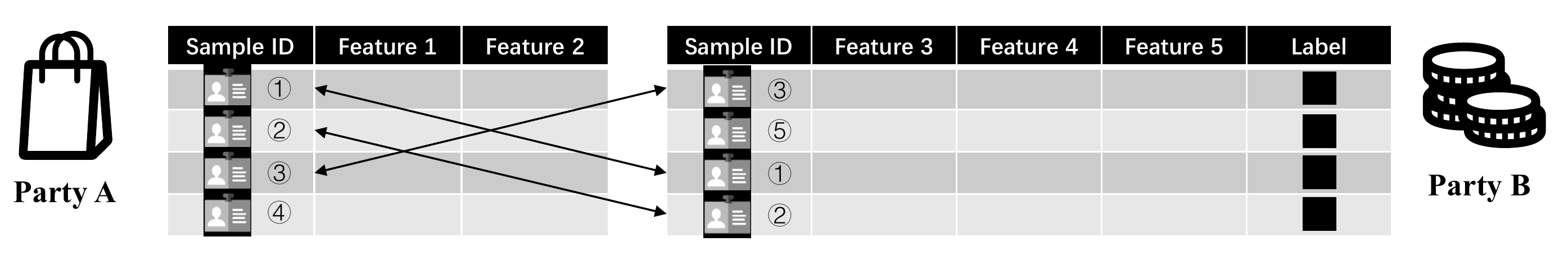}
    \caption{A typical scenario for using VFL involves Party A, an e-commerce company with features 1 and 2, and Party B, a bank with features 3, 4, 5 and the label. Together, they train a model that predicts loan approval decisions.}
    \label{fig:vfl}
\end{figure}

Vertical federated learning (VFL) \cite{xu2020federated,hard2018federated,yang2019federated,brisimi2018federated} is a trending paradigm that addresses a common dilemma faced by companies that share the same user group but differ in the features.
The concept is illustrated in Fig.~\ref{fig:vfl}. 
Suppose a bank, Party B, needs to improve its loan approval prediction and requires additional information from an e-commerce company, Party A.
In VFL, instead of directly exchanging user data, each party uploads the intermediate results of their user features calculated by a bottom model to a neutral party for further processing. 
This way, the raw data remains confidential.

The main challenge in VFL is ensuring these intermediate results' privacy and security. 
Current frameworks adopt Secure Multi-Party Computation (SMC) methods, such as Homomorphic Encryption (HE) \cite{hardy2017private,zhang2020batchcrypt}, to provide these guarantees. 
HE allows computations to be performed in an encrypted environment, ensuring no one can access the intermediate results in plain text.

However, recent research has shown that such methods are insufficient. 
In particular, the studies in \cite{luo2020feature,weng2021privacy,qiu2022relation,fu2022label} have demonstrated that an adversary can reconstruct the intermediate results, and even the raw data of the target party by using the sample's posteriors and the parameters of the VFL model. 
The reason behind these successful data reconstruction attacks is the ability of deep neural networks (DNNs) to model the correlation between the intermediate calculations and the raw inputs. 
For instance, generative adversarial networks (GANs) \cite{Creswell2018gan,Karras2020stylegan} have achieved remarkable success in image reconstruction.
Researchers have investigated several ways \cite{sun2021defend,Vepakomma2020NoPeekIL} to defend against data reconstruction attacks in VFL, such as increasing the correlation distance between the input and the corresponding activations.
These methods do reduce information leakage, but not completely. 
In the worst case where the adversary knows the model and the output for a target sample \cite{He2019ModelIA}, there remains a chance of the reconstruction attack. 

To eliminate the reversibility, we propose a new VFL framework called HashVFL that uses hashing.
The one-way nature of hashing allows our framework to block all attempts to recover intermediate calculations or raw inputs from the hash codes.
However, the integration of hashing makes it difficult for models to learn during training because the gradients disappear.
Moreover, hashing discards information, and to preserve privacy, the length of the hash code must be minimized. 
These two factors will inevitably affect the model's performance.
Given the above considerations, the design of our HashVFL framework addresses the following three challenges:
\begin{itemize}
    \item \textbf{Learnability.} The challenge lies in balancing the trade-off between preserving privacy through hashing and ensuring the learnability of the model. 
    The solution is to identify hash functions with easily estimable gradients so that the model can continue to train and maintain its performance.
    
    \textit{Our solution:} To achieve the desired trade-off between privacy preservation and learnability, we use the following approach: 
    1) We add a $Sign$ function to binarize the intermediate calculations of each party. 
    This is a common technique used in hashing \cite{su2018greedyhash} and binary neural networks \cite{Qin2020BinaryNN}. 
    2) We employ the \textit{Straight-Through Estimator} \cite{Bengio2013EstimatingOP,yin2018understanding} in back-propagation. 
    This allows the gradient to pass through the $Sign$ function exactly as it is, avoiding gradient vanishing.
    
    \item \textbf{Bit balance.} Hashing leads to the loss of information in each bit.
    Furthermore, to minimize the risk of information leakage, it is desired to limit the length of the hash codes.
    To address this challenge, we introduce the concept of \textit{bit balance}.
    This refers to maximizing the information carried by each bit, given a limited hash code length. 
    Ideally, we aim for half of the samples to take a value of 1/-1 on each bit. 
    This maximizes the information that the whole hash code can carry.
    
    \textit{Our solution:} To address the above requirement, we propose the use of \textit{Batch Normalization} (BN) \cite{sergey2015batchnorm}. 
    BN normalizes the intermediate calculations of each dimension in a batch of samples to have a standard normal distribution.
    It means that roughly half of the samples in a batch will have positive values, and half will have negative values at each dimension.
    Hence, incorporating BN helps address the issue of the effectiveness of the bit.
    
    \item \textbf{Consistency.} Since the binarization maps intermediate results into the same latent space, intuitively, a sample's hash codes from different parties should be consistent.
    One way to do this is to add the constraint in training by comparing the difference of hash codes between parties.
    However, this approach may result in a high computational overhead if there are many parties, which will definitely limit the application of VFL.
    
    \textit{Our solution:} To address the high computational overhead when comparing the hash codes of samples across many parties, a solution is to pre-define a set of binary codes for each class \cite{fan2020dpn}.
    This way, each party only needs to compare their hash codes with these binary codes, reducing the complexity from $O(N^2)$ to $O(N)$, where $N$ is the number of parties, making it suitable for scenarios with many parties.
    Additionally, the calculated differences between the sample's and target binary codes can also guide the optimization, as shown in \cite{hoe2021one}.
\end{itemize}

Experimental results demonstrate that our proposed HashVFL maintains the performance of the main task across various data types. 
Furthermore, HashVFL provides additional advantages by reducing label leakage, mitigating adversarial attacks, and detecting abnormal inputs.
Additionally, we assess HashVFL's performance in different conditions, such as varying numbers of parties, and find that it effectively handles various scenarios.

Our contributions are as follows:
\begin{itemize}
    \item We propose a novel approach for enhancing data security and privacy in the VFL by integrating hashing techniques.
    \item We address three key challenges in the design of the hashing-based VFL framework, i.e., learnability, bit balance, and consistency, and present a practical solution.
    \item We conduct a comprehensive empirical evaluation of our framework, HashVFL, demonstrating its ease of use, versatility, and effectiveness in defending against data reconstruction attacks. 
\end{itemize}

\section{Background}
\label{sec:background}
This section serves as an introduction to the background information relevant to VFL, data reconstruction attacks, hashing, and our threat model.
TABLE~\ref{table:notations} summarizes the necessary notations in this paper.

\begin{table}[t]
\centering
\caption{Symbols and notations.}
\label{table:notations}
\small
\resizebox{\columnwidth}{!}{%
\begin{tabular}{@{}r|p{8cm}@{}}
\multicolumn{1}{l|}{Notations} & Definition                            \\\midrule \midrule 
$P_i$, $D_i$                       & the $i$-th party in VFL, and $P_i$'s dataset \\
$\mathcal{U}_i$, $\mathcal{F}_i$   & $D_i$'s sample/user space and feature space            \\
$f_i$, $f_{top}$         &   $P_i$'s bottom model and the top model            \\
$\theta_i$, $\theta_{top}$          &    $f_i$'s parameters, and $f_{top}$'s parameters             \\
$\textbf{x}_i^{(u)}$, $\textbf{v}_i^{(u)}$, $\textbf{h}_i^{(u)}$          &  a sample $u$'s feature vector of $D_i$, $\textbf{x}_i^{(u)}$'s corresponding output from $f_i$, and the hash code of $\textbf{v}_i^{(u)}$            
\end{tabular}%
}
\end{table}

\subsection{Vertical Federated Learning}
Consider a set of $N$ parties $\left\{P_1,P_2,\cdots,P_N\right\}$ working on a classification task, each having its own dataset $\left\{D_1,D_2,\cdots,D_N\right\}$. Each dataset $D_i$ can be described as $(\mathcal{U}_i,\mathcal{F}_i)$, where $\mathcal{U}_i$ is the sample/user space and $\mathcal{F}_i$ is the feature space.

Before training, the parties must establish an overlapped sample space $\mathcal{U}$ as the intersection of all the sample spaces, i.e., $\mathcal{U}=\bigcap_{i=1}^N\mathcal{U}_i$. Then, the features of the samples in $\mathcal{U}$ from different $\mathcal{F}_i$ will be aligned based on their new indices in $\mathcal{U}$.

After the preparation of training set, $P_i$ trains its bottom model, $f_i$.
Let $\textbf{x}_i^{(u)}$ denotes the feature vector (raw input) of the sample $u$ from $\mathcal{F}_i$.
The function of $f_i$ is to map it into a $\Tilde{d}$-dimensional latent space, i.e., $f_i(\textbf{x}_i;\theta_i):\mathbb{R}^{d_i}\to\mathbb{R}^{\Tilde{d}}$, where $\theta$ denotes the model's parameters, and $d_i$ refers to the size of $\textbf{x}_i$'s dimension.
We use $\textbf{v}_i^{(u)}$ to represent $u$'s output of $f_i$.

Then, $\textbf{v}_i^{(u)}$ will be sent to a neutral third party's server for aggregation and further calculation.
Specifically, let $\textbf{v}_{cat}^{(u)}=[\textbf{v}_1^{(u)},\textbf{v}_2^{(u)},\cdots,\textbf{v}_N^{(u)}]$ denote the concatenated vector of the sample $u$ and $f_{top}$ denote the top model deployed at the server.
$f_{top}$ is supposed to learn a mapping from $\textbf{v}_{cat}^{(u)}$ to $\textbf{v}_{top}^{(u)}$, where $\textbf{v}_{top}^{(u)}$ is also the posterior for classification.
Formally, $f_{top}$ can be presented as $f_{top}(\textbf{v}_{cat};\theta_{top}):\mathbb{R}^{N\times\Tilde{d}}\to\mathbb{R}^{C}$, where $C$ denote the number of classes.

Finally, $\textbf{v}_{top}^{(u)}$ will be sent to the party who owns the label.
Then, the party calculates the loss, e.g., cross-entropy loss, and the corresponding gradients.
Using the chain rule, each model's parameters can be updated by passing the gradients.
The above process can be formulated as:
$$
    \min_{\left\{\theta_i\right\}_{i=1}^{N},\theta_{top}} \mathbb{E}_{u\in\mathcal{U}}
    [\ell(\textbf{x}_1^{(u)},\textbf{x}_2^{(u)},\cdots,\textbf{x}_N^{(u)},y;\left\{\theta_i\right\}_{i=1}^{N},\theta_{top})],
$$
where $\ell$ denotes the loss function, and $y$ is the label of $u$.

During the training process, intermediate calculations and gradients are transmitted, which are confidential information. To ensure privacy protection, existing frameworks, such as FATE \cite{ref:fate}, PySyft \cite{ref:pysyft}, TF Encrypted \cite{ref:tfencrypted}, and CrypTen \cite{ref:crypten}, employ homomorphic encryption (HE) \cite{gentry2009fully}. 
HE allows for vector operations, such as addition and multiplication, to be performed on encrypted data. 
In a nutshell, HE provides a secure environment for mathematical operations on sensitive information.

\subsection{Data Reconstruction Attack}
One persistent criticism of deep neural networks (DNNs) is their potential to violate user privacy by leaking information through the collected data. This issue is particularly prevalent in computer vision tasks that involve images.

In \cite{Zhu2019DeepLF}, Zhu et al. demonstrated that sensitive information can be reconstructed from leaked gradients. They achieved this by allowing the gradients of generated samples to closely approximate the gradients of target samples, resulting in high performance.
The attack can be formulated as follows:
$$
    \Tilde{\textbf{x}}^{*},\Tilde{y}^{*}=\mathop{\arg\min}_{\Tilde{\textbf{x}},\Tilde{y}}=\left\|\frac{\partial \ell(f(\Tilde{\textbf{x}},\theta),\Tilde{y})}{\partial \theta}-\nabla \theta\right\|^2,
$$
where $\Tilde{\textbf{x}}^{*}$, $\Tilde{y}^{*}$ are the reconstructed sample and its inferred label; $\ell(\cdot)$ denotes the loss function; $f$, $\theta$ denote the model and its parameters; and $\nabla \theta$ is the target sample's gradients.

In \cite{Pasquini2021UnleashingTT}, Pasquini et al. showed that an adversary could reconstruct the target image with knowledge of the model and the target image's posteriors under split learning scenarios.
Ergodan et al. \cite{Erdogan2021UnSplitDM} showed that the knowledge can further be relaxed to the model structure's copy and the target sample's intermediate calculations.
In \cite{He2019ModelIA}, He et al. unified model stealing and data reconstruction attack.
The attack can be formulated as follows:
$$
\left\{
    \begin{array}{ll}
         \Tilde{\textbf{x}}^*=\mathop{\arg\min}_{\Tilde{\textbf{x}}} \ell(f_{\Tilde{\theta}}(\Tilde{\textbf{x}}),f_{\theta}(\textbf{x}))+L(\Tilde{\textbf{x}}) \\
         \Tilde{\theta}^*=\mathop{\arg\min}_{\Tilde{\theta}} \ell(f_{\Tilde{\theta}}(\Tilde{\textbf{x}}),f_{\theta}(\textbf{x})),
    \end{array}
\right.
$$
where $\Tilde{\textbf{x}}^*$ is the reconstructed sample; $f$, $\theta$ denote the model and its parameters; $\Tilde{\theta}^*$ is the approximated parameters; $\ell(\cdot,\cdot)$ measures the distance between two terms; and $L(\Tilde{\textbf{x}})$ denotes the penalty function of $\Tilde{\textbf{x}}$ to guide the generation.
For example, in \cite{He2019ModelIA}, they used the Total Variation term \cite{Rudin1992NonlinearTV} to smooth the noise.
The optimization was alternated between samples and parameters to achieve the best results.

In \cite{luo2020feature}, Luo et al. revealed that the encryption mechanism in VFL cannot prevent the adversary from reconstructing the data by a generative adversarial network.
Weng et al. \cite{weng2021privacy} also verified the conclusion across more machine learning algorithms.
Moreover, in \cite{qiu2022relation}, Qiu et al. also showed that the reconstructed intermediate calculations could reflect the topology information used in graph neural networks.

In summary, side-channel information, such as gradients and intermediate calculations, can potentially reveal sensitive information due to the approximation capabilities of DNNs.

\subsection{Hashing}
Conventional hash functions, such as MD5 \cite{Rivest1990TheMM}, are data-independent, meaning they do not retain information about the input data. 
They take an input of arbitrary length and produce a fixed-length output, commonly referred to as a `fingerprint' or `message digest', through various mathematical operations.

In contrast to conventional hash functions, data-dependent hash methods retain information about the input data in their design. 
This is required for tasks such as similar image retrieval or product recommendations.
An example of a data-dependent hash method is Locality-Sensitive Hashing (LSH) which is widely used for Approximate Nearest Neighbor (ANN) search. 

Specifically, for two samples $u$ and $v$, LSH requires that their hash codes should have the property:
$$
    Pr[h(\textbf{x}^{(u)})=h(\textbf{x}^{(v)})]:\left\{
    \begin{array}{cll}
        \geq p_1 & if\,   & d(\textbf{x}^{(u)},\textbf{x}^{(v)}) \leq d_1\\
        \leq p_2 & else\, & d(\textbf{x}^{(u)},\textbf{x}^{(v)}) \geq d_2,
    \end{array}
    \right.
$$
where $\textbf{x}^{(u)}$ denotes the sample $u$'s feature vector, $h(\cdot)$ is the hash function, $d(\cdot)$ is the distance calculation function, $p_1$, $p_2$, and $d_1$, $d_2$ are the specific values of probability and distance.
The property means that for two samples' feature vectors, if their distance is less than $d_1$, their hash code should at least have the probability $p_1$ to have the same value; on the contrary, if it is less than $d_2$, the probability of their hash codes are same should not beyond $p_2$.

Recently, there has been a growing body of work \cite{Wu2019hashing,Wang2018hashing,su2018greedyhash,Cao2017HashNetDL} that has demonstrated the ability of DNNs to maintain the data-dependent property of hashing for approximate nearest neighbor search. These methods extract abstract representations of the data using DNNs and then binarize the representations in order to retain the correlation between the samples and maintain the effectiveness of retrieval. 
Our HashVFL also leverages these works to address learnability.

\subsection{Threat Model}
\label{sec:tm}
The threat model is based on the assumption of honest-but-curious adversaries \cite{Paverd2014ModellingAA}, meaning that all parties and the server will follow the requirements specified in VFL but may try to learn information about the intermediate calculations and raw data located on the target party. 

Additionally, it is assumed that the adversary knows each other's bottom model and samples' hash codes, but the local data of each party is strictly confidential. 
This is the strongest assumption for data reconstruction attacks and if HashVFL can defend against attacks under these conditions, it is likely to be effective with weaker assumptions where the adversary has less knowledge.
\section{Methodology}
\label{sec:method}
This section provides the framework of HashVFL and the implementation details of each component.

\subsection{Overview of HashVFL}
\begin{figure}[t]
    \centering
    \includegraphics[width=\columnwidth]{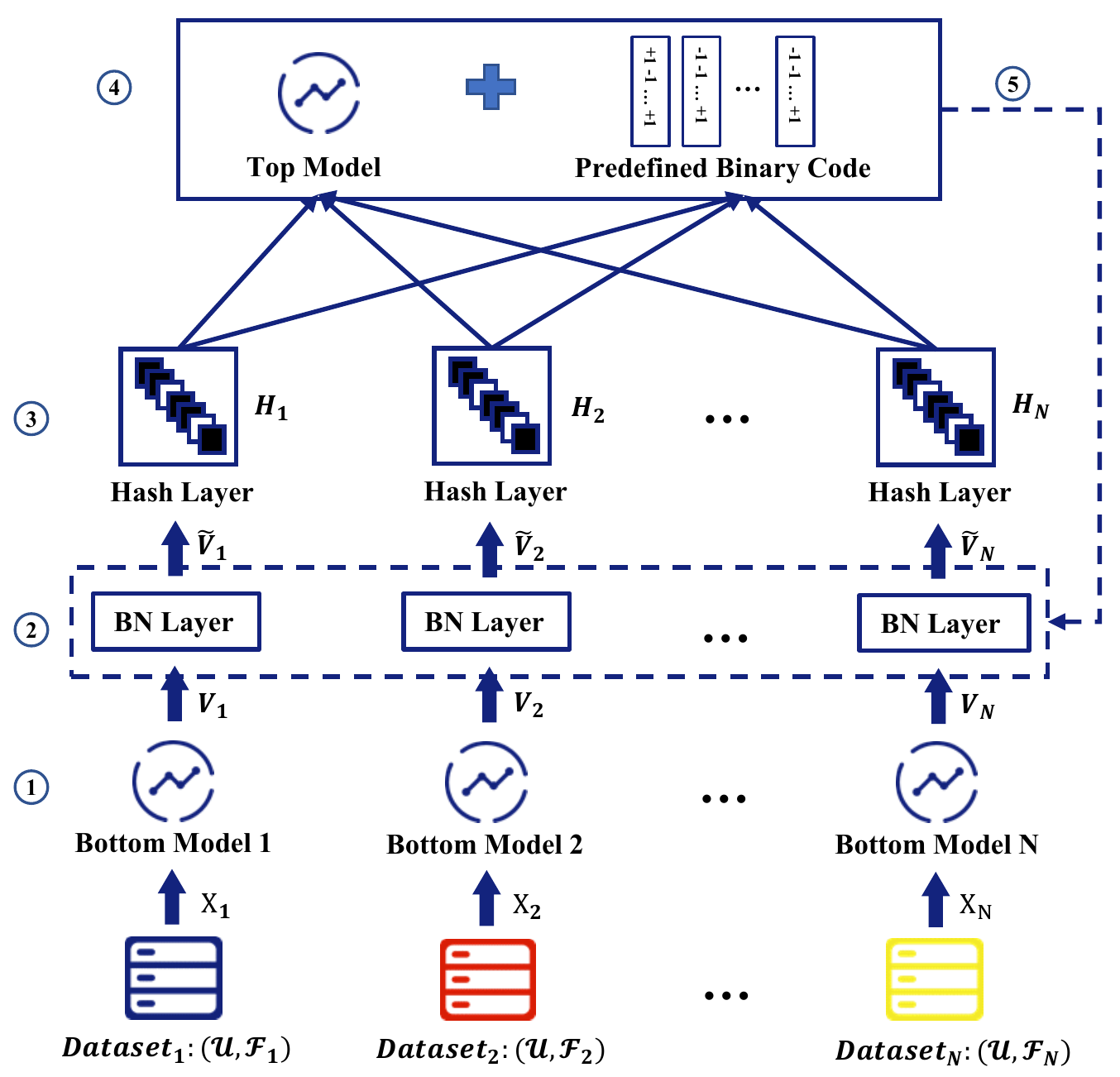}
    \caption{Overview of HashVFL. 1) Each party uses its bottom model to extract abstractions from local data. 2) The extracted abstractions are then normalized through a BN layer. 3) Normalized abstractions obtained from 2) are binarized by the hash layer and then uploaded to the server. 4) The server uses the top model to calculate classification loss and the distance between these codes and their target pre-defined binary codes. 5) The server calculates the gradients and transmits them back to corresponding parties. Please note that the gradients pass through the hash layer as they are due to the utilization of the Straight-Through Estimator (STE), enabling updates to commence from the batch normalization layer.}
    \label{fig:hvfl}
\end{figure}

Figure~\ref{fig:hvfl} illustrates the pipeline of HashVFL.
Following is a summary of the details.

First, each party, $P_i$, where $i\in\left\{1,2,\cdots,N\right\}$ and $N$ is the number of parties, prepares the training dataset $D_i=(\mathcal{U},\mathcal{F}_i)$.
Then, $P_i$ selects a specific model, $f_i$, for extracting information from the raw data.
For example, if $D_i$ is an image dataset, ResNet \cite{He2016ResNet} and VGG \cite{Simonyan2015VeryDC} are feasible candidates, which are popular model architectures for image classification.
$\textbf{x}_i$ denotes a sample's feature vector from $D_i$, and $\textbf{X}_i$ denotes a batch of samples' feature vectors.

Next, the outputs of $\textbf{X}_i$ from $f_i$, i.e., $\textbf{V}_i$, go through a Batch Normalizing transform (BN) layer, which is mandatory.
The transformed outputs, $\Tilde{\textbf{V}}_i$, now achieve a balance at each dimension, thus satisfying our `bit balance' requirement.
Then, these balanced outputs have to be binarized by a hash layer.
$\textbf{H}_i$ denotes the hash codes.

Finally, $\textbf{H}_i$ are concatenated at the server side, i.e., $\textbf{H}=[\textbf{H}_1,\cdots,\textbf{H}_N]$.
The top model, $f_{top}$, then calculates the posteriors of $\textbf{H}$.
Then, the posteriors and their ground truth $\textbf{Y}$ will be compared to calculate classification loss, usually cross-entropy (CE) loss.
Meanwhile, $\textbf{H}_i$ is also compared to pre-defined binary codes $\textbf{o}\in\left\{-1,+1\right\}^{C\times\Tilde{d}}$, where $C$ is the number of classes and $\Tilde{d}$ is the size of the hash code, for the consistency requirement.
Specifically, we calculate the distance between $\textbf{h}_i^{(u)}\in\textbf{H}_i$ and its target binary code $\textbf{o}_y$, where $u$ denotes a sample in the batch with label $y$, $\textbf{h}_i^{(u)}$ refers to $u$'s hash code, and $\textbf{o}_y$ denotes the class $y$'s corresponding code.

In summary, the loss function is designed as follows:
$$
    \mathcal{L}=CE(f_{top}(\textbf{H}),\textbf{Y})+(\textbf{1}-Cos(\textbf{H},\textbf{o}_{\textbf{Y}})),
$$
where $CE(\cdot,\cdot)$ denotes the CE loss term, $(\textbf{1}-Cos(\cdot,\cdot))$ denotes the cosine distance loss term, and $\textbf{o}_\textbf{Y}$ denotes $\textbf{Y}$'s corresponding pre-defined codes.

During the back-propagation process, the gradients are passed as they are through the hash layer.
Therefore, only BN layers' parameters and $\theta_i$ need to be updated.
Algorithm~\ref{alg:hvfl} describes the mainframe of HashVFL in training.

\begin{algorithm}
    \caption{Mainframe of HashVFL in training.}
    \label{alg:hvfl}
    \begin{algorithmic}[1]
        \REQUIRE $\left\{D_i,f_i\right\}_{i=1}^N$, $f_{top}$, pre-defined binary codes $\textbf{o}$
        \ENSURE $\left\{f_i\right\}_{i=1}^N$, $f_{top}$ for inference
        \FOR{each epoch}
            \FOR{each batch $(\textbf{X},\textbf{Y})$}
                \STATE \textbf{During forward process:} \
                \FOR{At each $P_i$}
                    \STATE $\textbf{V}_i \gets f_i(\textbf{X}_i)$ \
                    \STATE $\Tilde{\textbf{V}}_i \gets BN(\textbf{V}_i)$ \
                    \STATE $\textbf{H}_i \gets Sign(\Tilde{\textbf{V}}_i)$ \
                    \STATE Send $\textbf{H}_i$ to the server \
                \ENDFOR
                \STATE At the server:
                \STATE $\textbf{H} \gets concate(\left\{\textbf{H}_i\right\}_{i=1}^N)$ \
                \STATE $\textbf{o}_\textbf{Y} \gets onehot(\textbf{Y}) \times \textbf{o}$ \
                \STATE $\mathcal{L} \gets CE(f_{top}(\textbf{H}),\textbf{Y}) + (\textbf{1}-Cos(\textbf{H},\textbf{o}_\textbf{Y}))$ \
                \STATE \textbf{During backward process:} \
                \STATE At the server:
                \FOR{each $\textbf{H}_i$}
                    \STATE Calculate $\frac{\partial \mathcal{L}}{\textbf{H}_i}$
                    \STATE Send $\frac{\partial \mathcal{L}}{\textbf{H}_i}$
                \ENDFOR
                \FOR{At each $P_i$}
                    \STATE $\frac{\partial \mathcal{L}}{\partial \Tilde{\textbf{V}}_i} \gets \frac{\partial \mathcal{L}}{\textbf{H}_i}$
                    \STATE Update the following parameters
                \ENDFOR
            \ENDFOR
        \ENDFOR
    \end{algorithmic}
\end{algorithm}

In the following, we present the details of our implementation of the BN layer, the Hash layer, and the design of pre-defined binary codes.

\subsection{BN Layer}
Batch Normalization (BN) was first proposed by Sergey et al. \cite{sergey2015batchnorm} to solve the problem of \textit{Internal Covariate Shift}.
That is, during neural networks' training, the distribution of activations will shift due to the change in networks' parameters, 
In this paper, however, we use the design of BN to address our proposed bit balance.

Formally, given a batch of samples 
$
\mathcal{B}=\left\{\textbf{x}^{(1)},\textbf{x}^{(2)},\cdots,\textbf{x}^{(m)}\right\},
$
where $m$ denotes the size of the batch, a BN layer first normalizes each $\textbf{x}^{(i)}$ with batch mean $\mu_{\mathcal{B}}:\frac{1}{m}\Sigma_{i=1}^m\textbf{x}^{(i)}$ and batch variance $\sigma_{\mathcal{B}}^{2}:\frac{1}{m}\Sigma_{i=1}^m(\textbf{x}^{(i)}-\mu_{\mathcal{B}})^2$, i.e., $\Bar{\textbf{x}}^{(i)}:\frac{\textbf{x}^{(i)}-\mu_{\mathcal{B}}}{\sqrt{\sigma^2_{\mathcal{B}}+\epsilon}}$, where $\Bar{\textbf{x}}^{(i)}$ denotes normalized value, and set $\epsilon$ to prevent from division by zero.
Hence, we have $\Sigma_{i=1}^m\Bar{\textbf{x}}^{(i)}=0$ and $\frac{1}{m} \Sigma_{i=1}^m \Bar{\textbf{x}}^{2(i)}=1$, if we neglect $\epsilon$.
In such a way, we can guarantee that in each batch, the samples are evenly assigned positive and negative values on each bit.

However, simply normalizing each layer's input may change what the layer can represent \cite{sergey2015batchnorm}.
Therefore, there are two more parameters $\gamma$ and $\beta$ in the BN layer to scale and shift the normalized value: $\Tilde{\textbf{x}}^{(i)}=\gamma\Bar{\textbf{x}}^{(i)}+\beta$.
This way, the BN layer can recover the original activations if that is optimal for training.

During inference, for a batch of samples $\mathcal{B}_{inf}$, we transform them with $\Tilde{\textbf{x}}=\frac{\gamma}{\sqrt{Var[\textbf{x}]+\epsilon}}\Bar{\textbf{x}}+(\beta-\frac{\gamma \mathbb{E}[\textbf{x}]}{\sqrt{Var[\textbf{x}]+\epsilon}})$, where $\mathbb{E}[\textbf{x}]=\mathbb{E}_{\mathcal{B}_{inf}}[\mu_{\mathcal{B}_{inf}}]$ and $Var[\textbf{x}]=\frac{m}{m-1}\mathbb{E}_{\mathcal{B}_{inf}}[\sigma^2_{\mathcal{B}_{inf}}]$.

\subsection{Hash Layer}
Learning to hash is an NP-hard binary optimization problem \cite{Weiss2008SpectralH}. 
Therefore, a line of work \cite{Cao2017HashNetDL} adopts $tanh$ or $sigmoid$ for approximation, which ensures the differentiability.
However, these attempts still leave the risks of leakage in training.
Therefore, we use the $Sign$ function for binarization, which provides protection from training to inference.
Formally, the $Sign$ function is defined as follows:
$$
    h=Sign(v)=\left\{\begin{array}{cl}
    +1 & if\,\space v \ge 0 \\
    -1 & otherwise,
    \end{array}
    \right.
$$
where $h$ is the binary value of the input $v$.
For vectors, the function operates element-wise.

To solve the vanishment of gradients by using $Sign$, we combine the \textit{Straight-Through Estimator} (STE) in back-propagation.
In \cite{Bengio2013EstimatingOP}, to solve the challenge of estimating the gradients when stochastic or hard non-linearity neurons are used in neural networks, Bengio et al. proposed four estimators.
STE is the most efficient solution, as it behaves like the identity function.
Specifically, given a vector $\textbf{v}$ and its hash code $\textbf{h}$, which is obtained through the $Sign$ function, the gradients $\textbf{g}$ of $\textbf{v}$ can be estimated as:
$$
    \textbf{g}=\frac{\partial \mathcal{L}}{\partial \textbf{v}}=\frac{\partial \mathcal{L}}{\partial \textbf{h}}\cdot\frac{\partial \textbf{h}}{\partial \textbf{v}}\approx \frac{\partial \mathcal{L}}{\partial \textbf{h}},
$$
where $\mathcal{L}$ is the calculated loss of $\textbf{v}$.
With such an approximation, the model's training can thus continue.

% Yin et al. in their work \cite{yin2018understanding} presented a thorough assessment of STE for training quantized activation neural networks. 
% For those interested in this topic, further reading of relevant studies is recommended.

\subsection{Generation of Pre-defined Binary Codes}
The pre-defined binary codes are used to reduce the computation complexity in keeping the hash codes' consistency.
Moreover, they are also crucial for the classification task.

According to \cite{Charikar2002SimilarityET}, the probability of two vectors $\textbf{v}_i$ and $\textbf{v}_j$ having the same hash code under a family of hash functions using random hyperplane techniques is $1-\frac{\theta_{ij}}{\pi}$, where $\theta_{ij}$ denotes the angle between $\textbf{v}_i$ and $\textbf{v}_j$.
Therefore, to make the sample's hash codes discriminative for the task, we should let the pre-defined binary codes be as independent as possible, i.e., orthogonal to each other.

To achieve the orthogonality, we follow the practice in \cite{hoe2021one}, i.e., randomly generating the binary codes according to the Bernoulli distribution with $p=\frac{1}{2}$, where $p$ denotes the probability of signing $+1$ on each bit.
The derivation of $p$'s value is as follows.

First, for two randomly generated binary codes $\textbf{o}_1$ and $\textbf{o}_2$ with size $n$, we have
$
    Pr(cos(\textbf{o}_1,\textbf{o}_2)=0)=Pr(\Sigma_{i=1}^n o_{1i}\cdot o_{2i}=0).
$
Since $o_{1i}$ and $o_{2i}$ both obey the Bernoulli distribution with $p$, the probability $q$ of that $o_{1i}$ and $o_{2i}$ have the same value is $p^2+(1-p)^2$.

Then, $\Sigma_{i=1}^n o_{1i}\cdot o_{2i}$ becomes the binomial distribution with $q$, i.e., $n$ consecutive Bernoulli trials.
Hence, $Pr(\Sigma_{i=1}^n o_{1i}\cdot o_{2i}=0)=\binom{n}{\frac{n}{2}}q^{\frac{n}{2}}(1-q)^{\frac{n}{2}}$.
For a specific $n$, $\binom{n}{\frac{n}{2}}$ is a constant.
With inequality $q(1-q) \leq (\frac{q+(1-q)}{2})^2$, where the equal sign is obtained when $q=1-q$, we have $q=\frac{1}{2}$ to maximize $Pr(cos(\textbf{o}_1,\textbf{o}_2)=0)$.
Therefore, $p^2+(1-p)^2=\frac{1}{2}$, where we finally prove that $p=\frac{1}{2}$ is the best.

\subsection{Metrics of Distance}
Hamming distance \cite{Norouzi2012HammingDM} is commonly used in binary codes' distance calculation, while in this paper, we mainly discuss cosine similarity.
The reason is that they are literally equal for binary codes \cite{hoe2021one}.

For example, given a sample's hash code $\textbf{h}$ and its corresponding target binary code $\textbf{o}$, the hamming distance between them is:
$
    H(\textbf{h},\textbf{o})=\frac{\Tilde{d}-\textbf{h}^T\textbf{o}}{2},
$
where $H(\cdot)$ is the hamming distance calculation function and $\Tilde{d}$ is the length of codes.

Then, since $\textbf{h}^T\textbf{o}=\left\|\textbf{h}\right\|\left\|\textbf{o}\right\|cos \theta$, where $\left\|\cdot\right\|$ is the Euclidean norm and $\theta$ is the angle between $\textbf{h}$ and $\textbf{o}$, and both $\left\|\textbf{h}\right\|$ and $\left\|\textbf{o}\right\|$ equal to $\sqrt{\Tilde{d}}$, we have:
$
    H(\textbf{h},\textbf{o})=\frac{\Tilde{d}-\Tilde{d}cos\theta}{2}=\frac{\Tilde{d}}{2}(1-cos\theta).
$
Therefore, minimizing the hamming distance equals to minimizing the angle between the two binary codes, which also means maximizing the cosine similarity between them.
\section{Experimental Setup}
In our evaluation of the HashVFL, we use various datasets, models, and training details to assess its performance. 
We primarily consider the two-party VFL scenario following \cite{luo2020feature,weng2021privacy,qiu2022relation,fu2022label}, as it is the most popular scenario in the industry due to the consideration of communication cost and computation overhead \cite{yang2019federated}. 
However, the framework can be easily extended to multi-party scenarios, and we also evaluate the impact of the number of parties in Section~\ref{sec:parties}. 
The details of the datasets, models, and training procedures are presented in the following.

\subsection{Datasets}
Real-world VFL datasets \cite{hard2018keyboard,webankvflcase1,webankvflcase2} are proprietary and cannot be publicly accessed.
Therefore, we choose to evaluate on public datasets instead.
Specifically, we pick up six datasets, including three image datasets, two tabular datasets, and one text dataset:
1) MNIST \cite{LeCun1998mnist} is the most popular benchmark for evaluation, which has a training set of 60,000 examples and a test set of 10,000 examples;
2) CIFAR10 \cite{krizhevsky2009cifar} is another public dataset for image classification, which consists of 60,000 images with 10 classes;
3) FER \cite{Barsoum2016fer} is used for facial expression recognition, which consists of a training set of 28,709 examples and a test set of 7178 examples;
4) Company Bankruptcy Prediction Dataset (denoted by CBPD) \cite{liang2016bankruptcy} was collected from the Taiwan Economic Journal for the years from 1999 to 2009, which consists of 6,819 instances with 95 attributes and 2 classes;
5) CRITEO \cite{Guo2017criteo} is used for Click-Through-Rate (CTR) prediction, which consists of 100,000 instances;
6) IMDb \cite{maas2011imdb} is widely used in text analysis, consisting of a training set of 25,000 reviews and a test set of 25,000 reviews.

We remove the categorical features in CBPD and CRITEO, as removing them helps improve the performance from the practices of Kaggle \footnote{https://www.kaggle.com} (a community hosting competitions on data science) and the results of our experiments.
Furthermore, since CBPD and CRITEO are quite imbalanced in different classes, we use the over-sampling method to balance the number of samples in each category. 
Then, we split the training and test dataset with a ratio of $7:3$.

To simulate the VFL scenario, we follow the method described in \cite{luo2020feature,qiu2022relation,fu2022label}. 
For image data, if two parties hold the same ratio of features, we split the features from the center to ensure each party has an equal number of pixel columns. 
The same was done for tabular data, but with a different number of attributes for each party based on the feature ratio. 
In the case of text data, the difference between parties was the length of sentences.

\subsection{Models}
In our evaluation, the bottom model for image processing is ResNet \cite{He2016ResNet}, for text processing is BERT \cite{Devlin2019BERT}, and for tabular data is MLP \cite{Tang2016ExtremeLM}. 
The ResNet and BERT are downloaded from PyTorch \footnote{https://pytorch.org} and Hugging Face \footnote{https://huggingface.co}, respectively. 
The output dimension of the models is modified through a single linear layer, and their parameters are fine-tuned. 
The top model is a simple MLP with 1 hidden layer used to calculate the posteriors of the aggregated hash codes.

\subsection{Hyperparameters}
We set the training epochs for 30 times and an Adam \cite{Kingma2015AdamAM} optimizer with a batch size of 256 for images and tabular data, and 8 for texts (due to the limitation of memory).
The Adam optimizer has the learning rate of $1e-3$, the weight decay of $5e-4$, and a momentum by default in PyTorch's implementation.
In addition, we shrink the learning rate by 10\% every 10 epochs.
We store the last epoch's model, which are then used to measure the main task's performance on the test set.

\subsection{Environment}
We implement the attacks in Python and conduct all experiments on a workstation equipped with AMD Ryzen 9 3950X and an NVIDIA GTX 3090 GPU card. 
We use PyTorch to implement the models used in the experiments, and pandas \footnote{https://pandas.pydata.org} and sklearn \footnote{https://scikit-learn.org/stable} for data pre-processing.

\section{Evaluation}\label{sec:evaluation}
This section evaluates the proposed framework, HashVFL, on specific tasks and its defensive performance against data reconstruction attacks. 
Additionally, the extra benefit of HashVFL in detecting abnormal inputs and the need for combining differential privacy (DP) \cite{Dwork2014DP,Abadi2016DeepDP} are discussed.

\subsection{Performance Evaluation}\label{sec:performance}
We evaluate the performance of the HashVFL model under strict conditions by setting the length of the hash code to just enough to cover the number of classes, as redundant bits may leak information inadvertently \cite{melis2019exploiting}.
It is achieved by setting the length of the hash code to $\lceil \log_2 C \rceil$, where $C$ is the number of classes and $\lceil\cdot\rceil$ is the ceiling function. 
For example, on CIFAR10, the length of the hash code is set to 4 bits ($2^4$) to cover 10 classes.

TABLE~\ref{table:performance} summarizes the performance of HashVFL on different datasets. 
The results on MNIST, CIFAR10, and IMDb, show that HashVFL maintains the performance compared to results without defense, with the largest loss of accuracy on the test set being 5.39\% on CIFAR10. 

The results obtained on CBPD and CRITEO datasets demonstrate that HashVFL can even improve performance, as evidenced by the accuracy of 72.94\% on CRITEO with defense compared to 70.72\% without defense.
We speculate that this improvement may be attributed to the nature of tabular data, where the sign of values can be more informative for classification compared to floating-point numbers \cite{grinsztajn2022tree}.
Specifically, in our experimental design, we set the length of the embedding as 1 for binary classification tasks.
This design choice makes the classification task more challenging since the server only receives two floating-point numbers as input.
However, when applying the hash code, it effectively forces the bottom model to map the floating-point number to a binary class (e.g., +1/-1) first, and then enables the server to make the final judgment based on the submitted codes.
We speculate that this operation transforms the problem into a voting scenario for the top model, making it easier to learn and contribute to the observed performance improvement.

In conclusion, although the length of the hash code is limited to a minimum value, HashVFL satisfies the requirement of maintaining main task performance and can be applied to various data types.

\begin{table}[t]
\small
\centering
\caption{Accuracy (\%) on test set of HashVFL on various datasets.}
\label{table:performance}
\resizebox{\columnwidth}{!}{%
\begin{tabular}{@{}l|c|c|c|c|c|c@{}}
\toprule
Dataset          & MNIST     & CIFAR10       & FER           & CBPD          & CRITEO        & IMDb      \\ \midrule
Without Defense  & 98.99     & 76.22         & 55.93         & 50.30         & 70.72         & 73.64     \\ \midrule
With Defense     & 97.75     & 70.83         & 51.11         & 69.34         & 72.94         & 69.72     \\ \bottomrule
\end{tabular}%
}
\end{table}

\subsection{Defending against Reconstruction Attacks}
\label{sec:reconstruction}
In this section, we assess the defensive capability of HashVFL against data reconstruction attacks. 
Our analysis is based on the threat model outlined in Section~\ref{sec:tm}, which assumes that the adversary has complete knowledge of the victim's bottom model and the target sample's hash code. 
We examine HashVFL's protection of privacy from three perspectives: 
\begin{itemize}
    \item Recovery of the intermediate result from its hash code;
    \item Recovery of the target sample from its hash code;
    \item Revealing common features among a group of samples sharing the same code.
\end{itemize}

\textbf{Reconstructing the intermediate result from its hash code is intractable.}
Previous works such as \cite{su2018greedyhash,Cao2017HashNetDL,Wu2019hashing} imposed a constraint on the embeddings to be close to their binary codes using a loss term of Euclidean distance, in addition to hashing. 
This constraint ideally results in the intermediate results being equal to the hash codes. 
However, our proposed design, HashVFL, does not impose such a constraint. Consequently, it is not possible to reconstruct a sample's intermediate results using its hash codes. 
This fundamental difference sets HashVFL apart from previous works.

\textbf{Reconstructing the target sample from its hash code is impossible.}
Since the intermediate results are not accessible, the semi-honest adversary can only obtain a sample's hash code.
Then, we further conclude that the adversary cannot reconstruct a specific target sample's raw data, even with complete knowledge. 

On the one hand, \textit{Sign} function discards a significant amount of information.
On the other hand, the strict limitation on the length of the hash code, preventing the assignment of a unique hash code to a target sample. 
These factors reinforce the privacy guarantee of HashVFL.

\textbf{The same code shared by a group of samples may leak common features.}
A shorter hash code leads to more hash collisions. 
In our case, samples belonging to the same class are expected to have the same hash code. 
This raises the question of whether the adversary can uncover common features among samples of the same class.

\subsubsection{Measurement of Label Leakage}
It is evident that if our model achieves perfect classification performance, the hash code will also accurately represent the corresponding label.
Therefore, there exists a correlation between performance and label leakage in VFL.
In particular, in \cite{fu2022label}, Fu et al. highlighted that an adversary possessing a bottom model could deduce a sample's label based on the local embedding, with an accuracy that is proportional to the quality of the bottom model.
Hence, we further investigate whether the combination of hashing can reduce the degree of label leakage.

Specifically, we compare the label inference accuracy between the original embeddings and the binarized embeddings on the selected datasets, following the methodology of the passive label inference attack (PLA) outlined in \cite{fu2022label}.
The default length of the hash code is set to 16. The summarized results are presented in TABLE~\ref{table:label}.

\begin{table*}[t]
\caption{Label leakage measurement with different architectures under PLA.}
\label{table:label}
\tiny
\resizebox{\textwidth}{!}{%
\begin{tabular}{@{}lllllll@{}}
\toprule
Dataset & CBPD          & CRITEO        & MNIST         & CIFAR10       & FER           & IMDb          \\ \midrule
Base    & 62.41$\pm$4.38 & 72.31$\pm$0.82 & 95.05$\pm$0.13 & 70.13$\pm$0.09 & 52.09$\pm$0.22 & 55.64$\pm$0.85 \\
Ours    & 63.17$\pm$0.43 & 69.17$\pm$0.10 & 85.24$\pm$0.06 & 69.93$\pm$0.07 & 51.20$\pm$0.10 & 50.48$\pm$0.24 \\ \bottomrule
\end{tabular}%
}
\end{table*}

The results demonstrate that the inference performance of PLA decreases after the binarization process.
This reduction in performance can be attributed to the loss of a significant amount of information during the binarization step.
For example, on the MNIST dataset, the accuracy decreases from 95.05\% to 85.24\%, which is the highest reduction observed among all the datasets evaluated.

In conclusion, the introduction of the hashing mechanism not only effectively mitigates the data reconstruction attack but also helps alleviate the degree of label leakage to a certain extent.

\subsubsection{Visual Results}
In addition to considering label leakage, it is essential to examine what other common features the hash code may potentially reveal.
To address this question, we assess the defense performance of HashVFL on the image datasets.
It is because the reconstructed images provides an intuitive measure of the level of information leakage, specifically looking for any blurred contours in the reconstruction results.

Our approach is based on the method described in \cite{He2019ModelIA}. 
In this work, He et al. proposed a novel attack that could precisely recover images in a black-box scenario, achieving state-of-the-art results. 
Our study relaxes their assumption by allowing the adversary to have knowledge of the bottom model, making the attack stronger than in the black-box scenario.
Specifically, we reconstruct the features according to the following formula:
$$
    \Tilde{\textbf{x}}^*=\mathop{\arg\min}_{\Tilde{\textbf{x}}} \mathop{MSE}(f_{\theta}(\Tilde{\textbf{x}}),\textbf{o}_y)+\lambda TV(\Tilde{\textbf{x}}),
$$
where $\Tilde{\textbf{x}}^*$ is the generated sample; $f_{\theta}(\cdot)$ denotes the bottom model; $MSE(\cdot)$ calculates two vectors' mean square error; $\textbf{o}_y$ denotes the target class $y$'s corresponding binary code; $TV(\cdot)$ denotes the Total Variation (TV) term \cite{Rudin1992NonlinearTV}; $\lambda$ is a coefficient to modify the weight of TV term.

The interpretation of this formula is simple. 
If the adversary is aware of the model and the output of the target sample, he/she can reconstruct the target sample by creating $\Tilde{\textbf{x}}$, whose output is similar to that of the target sample. 
Similar idea can be seen in \cite{Zhu2019DeepLF}, where Zhu et al. used the target sample's gradients instead of the output.
In our case, since the hash code is the only information that the adversary can access, the formula is a reasonable one to choose.

Since the generation process may produce noise, the TV term is introduced to smooth the output.
The TV term is calculated as follows:
$$
    TV(\textbf{x})=\Sigma_{i,j}\sqrt{\left\|\textbf{x}_{i+1,j}-\textbf{x}_{i,j}\right\|^2+\left\|\textbf{x}_{i,j+1}-\textbf{x}_{i,j}\right\|^2},
$$
where $i$ and $j$ denote the pixel indices.

\begin{figure}[t]
    \centering
    \includegraphics[width=\columnwidth]{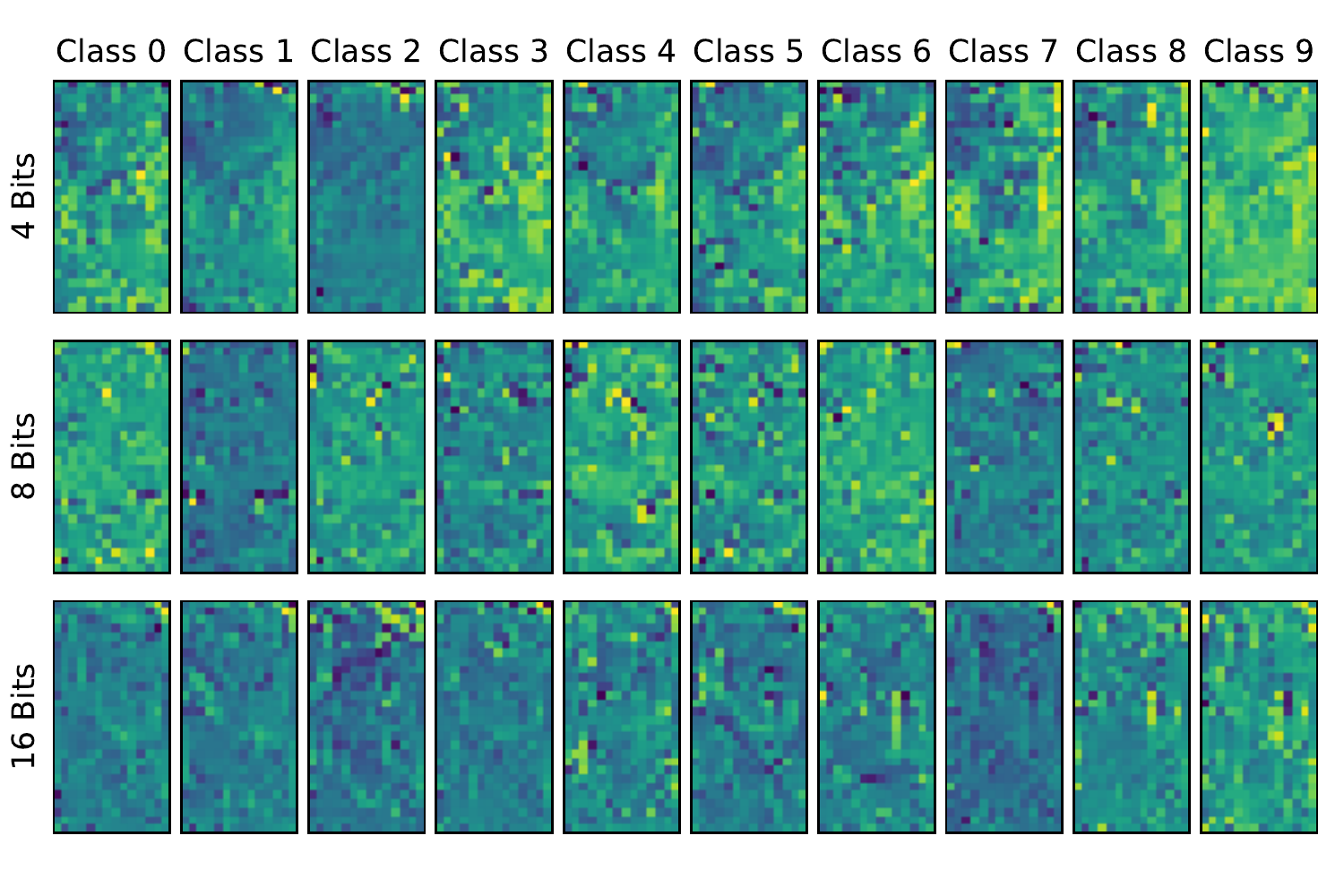}
    \caption{Revealing common features on MNIST. The first row shows the reconstructed results with 4 bits hash code on different classes. The second and third rows show 8 and 16 bits, respectively.}
    \label{fig:re-mnist}
\end{figure}

We conducted 3,000 rounds of reconstruction for each class code and varied the hash code length from 4 bits to 16 bits to examine the influence of code length on information leakage. 
The reconstructed results for MNIST are displayed in Fig.~\ref{fig:re-mnist}. 
The results demonstrate that even with the strongest assumption, the adversary cannot recover meaningful information. 
Moreover, the length of the code does not have a significant impact on the reconstruction process.

\subsubsection{Statistical Analysis}
In order to conduct a comprehensive analysis of the information leakage resulting from the reconstruction attack on the hash code, we employ three additional metrics:  Kullback–Leibler
divergence (KLD) \cite{Dong2017DroppingAO}, distance correlation (DCOR) \cite{Abuadbba2020CanWU}, and structural similarity index measure (SSIM) \cite{Vepakomma2020NoPeekIL}, as suggested by Pham et al. \cite{Pham2022BinarizingSL}.
These metrics enable us to provide a statistical evaluation of the leakage. The results are summarized in TABLE~\ref{table:stats}.

\begin{table}[t]
\caption{Statistical analysis of leakage with three different metrics.}
\label{table:stats}
\small
\resizebox{\columnwidth}{!}{%
\begin{tabular}{@{}l|lll@{}}
\toprule
\multirow{2}{*}{Dataset} & \multicolumn{3}{c}{Measure}                                                                            \\ \cmidrule(l){2-4} 
                         & \multicolumn{1}{c|}{KLD ($\ge 0$)}               & \multicolumn{1}{c|}{SSIM ($[0, 1]$)}          & \multicolumn{1}{c}{DCOR ($[0, 1]$)} \\ \midrule
MNIST                    & \multicolumn{1}{l|}{11.5757$\pm$7.2420}    & \multicolumn{1}{l|}{0.0157$\pm$0.0137} & 0.8757$\pm$0.0487            \\ \midrule
CIFAR10 & \multicolumn{1}{l|}{340.5571$\pm$270.2468} & \multicolumn{1}{l|}{0.0620$\pm$0.0366} & 0.8540$\pm$0.0916 \\ \midrule
FER                      & \multicolumn{1}{l|}{813.7469$\pm$536.4684} & \multicolumn{1}{l|}{0.0948$\pm$0.0511} & 0.9004$\pm$0.0693            \\ \bottomrule
\end{tabular}%
}
\end{table}

The results indicate that the KLD values are considerably large across all three datasets.
This means significant differences in the distributions of pixel values between the raw data and the reconstructed data.
Moreover, the SSIM scores for all three datasets are notably small, indicating a reduced semantic similarity between the reconstructed data and the raw data.
A similar trend is observed in the DCOR scores.

Based on these findings, we can conclude that the hashing operation in HashVFL effectively defends against the reconstruction attack and minimizes the leakage of common features.

\subsection{Defending against Adversarial Attack}
\label{sec:adv}
The purpose of this section is to investigate whether the employed hashing mechanism results in a loss of robustness in VFL models against adversarial attacks.
To thoroughly assess the robustness, we enhance the capabilities of the adversary based on the previous threat model, enabling them to fully execute adversarial attacks.
For example, we provide the adversary with complete knowledge of both the bottom models and the top model, allowing them to successfully carry out the projected gradient descent (PGD) attack \cite{Madry2017TowardsDL}.

PGD attack is one standard white-box adversarial attack widely used in the field.
Formally, it can be described as follows:
\begin{equation}
    \left\{
    \begin{aligned}
        \textbf{x}_m &= \textbf{x}_{m-1} - \eta * sign(\nabla_{\textbf{x}_{m-1}}\mathcal{L}(\textbf{x}_{m-1}, y_t; \theta)) \\
        \bm{\phi}_m &= clip(\textbf{x}_m-\textbf{x}_0, \omega) \\
        \textbf{x}_m &= \textbf{x}_0 + \bm{\phi}_m,
    \end{aligned}
\right.
\end{equation}
where $\textbf{x}_0$ denotes the original intermediate results, $\textbf{x}_m$ denotes the perturbed adversarial results at $m$-th optimization, $\eta$ is the step size, $y_t$ denotes the target class, and $clip(\textbf{x}_m-\textbf{x}_0, \omega)$ denotes the restriction that clips the perturbation $\bm{\phi}_m$ to a given threshold, which is $(-\omega, \omega)$.

To evaluate the performance of the proposed method, we select MNIST and CIFAR10 datasets for experimentation.
These two datasets are suitable for evaluating robustness against adversarial attacks as they both consist of 10 classes and are widely used benchmarks.
We conduct evaluations using different combinations of the threshold $\omega$ and step size $\eta$, while keeping the hash code fixed at 16 bits.
For the experiments, we randomly select 100 samples from all classes except class `0' and calculate the success rate of the attack when these samples are misclassified as class `0'.
The results are summarized in TABLE~\ref{table:adversarial}.

\begin{table}[t]
\caption{Defense performance of different framework with different hyperparameters on MNIST and CIFAR10.}
\label{table:adversarial}
\tiny
\resizebox{\columnwidth}{!}{%
\begin{tabular}{@{}l|l|l|ll@{}}
\toprule
\multirow{2}{*}{Dataset} & \multirow{2}{*}{Threshold} & \multirow{2}{*}{Step Size} & \multicolumn{2}{l}{Attack Success Rate} \\ \cmidrule(l){4-5} 
                         &                      &         & \multicolumn{1}{l|}{Base} & Hashing \\ \midrule
\multirow{4}{*}{MNIST}   & \multirow{2}{*}{$\omega=1$} & $\eta=0.1$ & \multicolumn{1}{l|}{85} & 0       \\ \cmidrule(l){3-5} 
                         &                      & $\eta=1$   & \multicolumn{1}{l|}{74} & 64    \\ \cmidrule(l){2-5} 
                         & \multirow{2}{*}{$\omega=2$} & $\eta=0.1$ & \multicolumn{1}{l|}{1}    & 0       \\ \cmidrule(l){3-5} 
                         &                      & $\eta=1$   & \multicolumn{1}{l|}{1}    & 1       \\ \midrule
\multirow{4}{*}{CIFAR10} & \multirow{2}{*}{$\omega=1$} & $\eta=0.1$ & \multicolumn{1}{l|}{76} & 0       \\ \cmidrule(l){3-5} 
                         &                      & $\eta=1$   & \multicolumn{1}{l|}{66} & 85    \\ \cmidrule(l){2-5} 
                         & \multirow{2}{*}{$\omega=2$} & $\eta=0.1$ & \multicolumn{1}{l|}{98} & 0       \\ \cmidrule(l){3-5} 
                         &                      & $\eta=1$   & \multicolumn{1}{l|}{95} & 64    \\ \bottomrule
\end{tabular}%
}
\end{table}

The results reveal that the hashing mechanism employed can actually enhance the robustness of VFL models.
This improvement can be attributed to the binarization operation, which requires that the values of the submitted codes from each party be limited to $\left\{-1, 1\right\}$.
This limitation, in turn, restricts the performance of the PGD attack when its step size and threshold are smaller than 1.
In other words, our hashing-based framework can expand the robust radius of each sample by at least 1 compared to the base VFL model without any defense mechanism.
Unfortunately, when the threshold and step size exceed 1, both frameworks fail to defend against the adversarial attack.

In conclusion, our HashVFL approach enhances the robustness of VFL models by forcing the embeddings to map to a fixed set of $\left\{-1, 1\right\}$, thereby expanding the robust radius of the samples. 

\subsection{Detecting Abnormal Inputs}
\label{sec:detection}
Adversarial attacks essentially exploit abnormal inputs to alter outcomes. Therefore, when we return to the multi-party computing framework of VFL, it means that the adversary's embedding should be inconsistent with other normal parties' embeddings.
From this intuition, we explore an additional advantage of HashVFL: its ability to efficiently detect abnormal inputs, as consistency requires identical code from each party. 
If the hash codes of one sample from two parties differ significantly, it may indicate abnormal inputs. 
For instance, if the hamming distance between the two codes is larger than half the length of the code, it may indicate cheating.

The detection capability of HashVFL is evaluated in a two-party scenario, where each party holds half the features. 
To verify our speculation, all combinations of different hash codes are detected. 
For each class, the code from one party, $P_{initiator}$, is set as the corresponding pre-defined binary code, and the other party's code, $P_{participant}$, is varied to observe differences between correct and incorrect predictions. Ideally, the hamming distance between the codes of correct predictions should be less than half the length, while it should be greater for incorrect predictions. 
This means that a malicious party must change at least half the bits to alter the prediction. 
The length of the hash code is 4 bits.

TABLE~\ref{table:detection4bits} summarizes the results.
In TABLE~\ref{table:detection4bits}, the `-' symbol denotes the absence of a wrong prediction, and `/' denotes the absence of the case in the dataset.
The results in the `Average' column confirm our speculation that if the hamming distance between two parties' codes is greater than half the length (2 in this case), the prediction is probably incorrect. 
This conclusion is also supported by most of the detailed results for each class. 
However, on CIFAR10, there are exceptions in the results for class `1' and class `2'.
These exceptions are due to a concentration of wrong predictions on specific classes and a relatively small hamming distance between the pre-defined codes of the classes.
For example, many wrong predictions on class `1' give class `5', and on class `2' give class `8'.
The pre-defined codes of class `1' and class `5' are $[-1,1,1,1]$ and $[1,1,1,1]$, whose hamming distance is relatively small.
Therefore, the average hamming distance of wrong predictions is smaller in such two cases.
The results on CBPD, CRITEO, and IMDb are consistent with the above conclusion. 

In summary, the HashVFL method efficiently detects abnormal inputs by computing the hamming distance between hash codes submitted by different parties. 
If the hamming distance is greater than half the length of the code, it suggests the possibility of cheating during inference.

\begin{table}[t]
\centering
\caption{Analysis of 4-bit hash code detection performance. The `Class' column shows the average hamming distance between the two hash codes. The `Average' column represents the average results for each class.}
\label{table:detection4bits}
\resizebox{\columnwidth}{!}{%
\begin{tabular}{@{}l|l|cccccccccc|c@{}}
\toprule
\multirow{2}{*}{Dataset} &
  \multirow{2}{*}{} &
  \multicolumn{10}{c|}{Class} &
  \multicolumn{1}{l}{\multirow{2}{*}{Average}} \\ \cmidrule(lr){3-12}
 &
   &
  \multicolumn{1}{c|}{0} &
  \multicolumn{1}{c|}{1} &
  \multicolumn{1}{c|}{2} &
  \multicolumn{1}{c|}{3} &
  \multicolumn{1}{c|}{4} &
  \multicolumn{1}{c|}{5} &
  \multicolumn{1}{c|}{6} &
  \multicolumn{1}{c|}{7} &
  \multicolumn{1}{c|}{8} &
  9 &
  \multicolumn{1}{l}{} \\ \midrule
\multirow{2}{*}{MNIST} &
  correct &
  \multicolumn{1}{c|}{0.67} &
  \multicolumn{1}{c|}{1.57} &
  \multicolumn{1}{c|}{1.50} &
  \multicolumn{1}{c|}{1.20} &
  \multicolumn{1}{c|}{1.00} &
  \multicolumn{1}{c|}{1.20} &
  \multicolumn{1}{c|}{1.17} &
  \multicolumn{1}{c|}{1.75} &
  \multicolumn{1}{c|}{1.33} &
  1.70 &
  1.31 \\ \cmidrule(l){2-13} 
 &
  error &
  \multicolumn{1}{c|}{2.31} &
  \multicolumn{1}{c|}{2.33} &
  \multicolumn{1}{c|}{2.50} &
  \multicolumn{1}{c|}{2.36} &
  \multicolumn{1}{c|}{2.23} &
  \multicolumn{1}{c|}{2.36} &
  \multicolumn{1}{c|}{2.50} &
  \multicolumn{1}{c|}{2.25} &
  \multicolumn{1}{c|}{2.40} &
  2.50 &
  2.38 \\ \midrule
\multirow{2}{*}{CIFAR10} &
  correct &
  \multicolumn{1}{c|}{1.00} &
  \multicolumn{1}{c|}{\textbf{2.10}} &
  \multicolumn{1}{c|}{\textbf{3.00}} &
  \multicolumn{1}{c|}{1.43} &
  \multicolumn{1}{c|}{1.00} &
  \multicolumn{1}{c|}{1.67} &
  \multicolumn{1}{c|}{1.43} &
  \multicolumn{1}{c|}{1.50} &
  \multicolumn{1}{c|}{1.20} &
  1.62 &
  1.59 \\ \cmidrule(l){2-13} 
 &
  error &
  \multicolumn{1}{c|}{2.33} &
  \multicolumn{1}{c|}{\textbf{1.83}} &
  \multicolumn{1}{c|}{\textbf{1.67}} &
  \multicolumn{1}{c|}{2.44} &
  \multicolumn{1}{c|}{2.33} &
  \multicolumn{1}{c|}{2.43} &
  \multicolumn{1}{c|}{2.44} &
  \multicolumn{1}{c|}{2.50} &
  \multicolumn{1}{c|}{2.36} &
  2.38 &
  2.27 \\ \midrule
\multirow{2}{*}{FER} &
  correct &
  \multicolumn{1}{c|}{\textbf{2.00}} &
  \multicolumn{1}{c|}{\textbf{2.00}} &
  \multicolumn{1}{c|}{1.90} &
  \multicolumn{1}{c|}{0.67} &
  \multicolumn{1}{c|}{1.29} &
  \multicolumn{1}{c|}{1.93} &
  \multicolumn{1}{c|}{1.00} &
  \multicolumn{1}{c|}{/} &
  \multicolumn{1}{c|}{/} &
  / &
  1.35 \\ \cmidrule(l){2-13} 
 &
  error &
  \multicolumn{1}{c|}{\textbf{2.00}} &
  \multicolumn{1}{c|}{\textbf{2.00}} &
  \multicolumn{1}{c|}{2.17} &
  \multicolumn{1}{c|}{2.31} &
  \multicolumn{1}{c|}{2.56} &
  \multicolumn{1}{c|}{2.50} &
  \multicolumn{1}{c|}{2.33} &
  \multicolumn{1}{c|}{/} &
  \multicolumn{1}{c|}{/} &
  / &
  2.27 \\ \midrule
\multirow{2}{*}{CBPD} &
  correct &
  \multicolumn{1}{c|}{1.87} &
  \multicolumn{1}{c|}{\textbf{2.00}} &
  \multicolumn{1}{c|}{/} &
  \multicolumn{1}{c|}{/} &
  \multicolumn{1}{c|}{/} &
  \multicolumn{1}{c|}{/} &
  \multicolumn{1}{c|}{/} &
  \multicolumn{1}{c|}{/} &
  \multicolumn{1}{c|}{/} &
  / &
  1.94 \\ \cmidrule(l){2-13} 
 &
  error &
  \multicolumn{1}{c|}{\textbf{4.00}} &
  \multicolumn{1}{c|}{-} &
  \multicolumn{1}{c|}{/} &
  \multicolumn{1}{c|}{/} &
  \multicolumn{1}{c|}{/} &
  \multicolumn{1}{c|}{/} &
  \multicolumn{1}{c|}{/} &
  \multicolumn{1}{c|}{/} &
  \multicolumn{1}{c|}{/} &
  / &
  4.00 \\ \midrule
\multirow{2}{*}{CRITEO} &
  correct &
  \multicolumn{1}{c|}{1.87} &
  \multicolumn{1}{c|}{\textbf{2.00}} &
  \multicolumn{1}{c|}{/} &
  \multicolumn{1}{c|}{/} &
  \multicolumn{1}{c|}{/} &
  \multicolumn{1}{c|}{/} &
  \multicolumn{1}{c|}{/} &
  \multicolumn{1}{c|}{/} &
  \multicolumn{1}{c|}{/} &
  / &
  1.94 \\ \cmidrule(l){2-13} 
 &
  error &
  \multicolumn{1}{c|}{\textbf{4.00}} &
  \multicolumn{1}{c|}{-} &
  \multicolumn{1}{c|}{/} &
  \multicolumn{1}{c|}{/} &
  \multicolumn{1}{c|}{/} &
  \multicolumn{1}{c|}{/} &
  \multicolumn{1}{c|}{/} &
  \multicolumn{1}{c|}{/} &
  \multicolumn{1}{c|}{/} &
  / &
  4.00 \\ \midrule
\multirow{2}{*}{IMDb} &
  correct &
  \multicolumn{1}{c|}{1.87} &
  \multicolumn{1}{c|}{\textbf{2.00}} &
  \multicolumn{1}{c|}{/} &
  \multicolumn{1}{c|}{/} &
  \multicolumn{1}{c|}{/} &
  \multicolumn{1}{c|}{/} &
  \multicolumn{1}{c|}{/} &
  \multicolumn{1}{c|}{/} &
  \multicolumn{1}{c|}{/} &
  / &
  1.94 \\ \cmidrule(l){2-13} 
 &
  error &
  \multicolumn{1}{c|}{\textbf{4.00}} &
  \multicolumn{1}{c|}{-} &
  \multicolumn{1}{c|}{/} &
  \multicolumn{1}{c|}{/} &
  \multicolumn{1}{c|}{/} &
  \multicolumn{1}{c|}{/} &
  \multicolumn{1}{c|}{/} &
  \multicolumn{1}{c|}{/} &
  \multicolumn{1}{c|}{/} &
  / &
  4.00 \\ \bottomrule
\end{tabular}%
}
\end{table}

\subsection{Analysis of Combining Differential Privacy}\label{sec:dp}
This section explores the need for further incorporating differential privacy (DP) \cite{Abadi2016DeepDP,Dwork2014DP} into our existing scheme. 
DP is a widely used privacy-enhancing technique in DNNs and has been integrated into frameworks such as FATE and TF Encrypted for data protection.

\paragraph{Theoretical Analysis}
In \cite{Pham2022BinarizingSL}, Pham et al. proposed a method to integrate DP with binary code:
$$
    \textbf{h}=Sign(Sign(f(\textbf{x}))+Lap(\frac{s}{\epsilon})),
$$
where $\textbf{h}$ is the hash code; $\textbf{x}$ is the feature vector; $f$ is a DNN model; $s$ is the sensitivity of $Sign(\cdot)$, actually 2 for binary codes; $\epsilon$ is the privacy budget \cite{Dwork2014DP}; and $Lap(\cdot)$ is the Laplace distribution sampling function.

From this design, if $|Lap(\frac{2}{\epsilon})|<1$, then $\textbf{h}=Sign(f(\textbf{x}))$.
The probability of $|Lap(\frac{2}{\epsilon})|<1$ can be calculated as:
$$
    Pr[|Lap(\frac{2}{\epsilon})|<1] = 1-[cdf(1)-cdf(-1)] = 1-e^{-\frac{\epsilon}{2}},
$$
where $cdf(\cdot)$ denotes the cumulative distribution function of Laplace.
According to the definition of Approximate DP \cite{dwork2006our}, the analysis also indicates that the hashing operation satisfies $(\epsilon, \delta)$-DP, where $\delta = 1-e^{-\frac{\epsilon}{2}}$.

However, the derivation also reveals a problem that the added noise cannot change the value of one bit when the privacy budget $\epsilon$ is large (indicating a weak privacy protection level).
Specifically, following the above calculation, we have $Pr[|Lap(\frac{2}{\epsilon})|\ge 1]=e^{-\frac{\epsilon}{2}}$.
Consider that there is half chance that the sign of the noise is the same as the bit, the probability of flipping one bit's sign is $\frac{1}{2}e^{-\frac{\epsilon}{2}}$.
When we set $\epsilon=10$, the probability decreases rapidly to 0.33\%, which almost does not provide any privacy protection.
In such a situation, the adversary can use the received hash code as the real value, and there is almost no error.

The analysis suggests that integrating DP in HashVFL is not necessary, if there is a large privacy budget.
In addition, the added noise may cause the performance of the main task to degrade.
Therefore, we do not recommend integrating DP in HashVFL as it has a limited defensive effect.

\paragraph{Experimental Demonstration}
To better understand the necessity of combining DP in HashVFL, we conduct experiments with different privacy budgets to evaluate its impact on main task's performance.
We still evaluated the two-party scenario, where each party holds half of the features. 
The results are summarized in TABLE~\ref{table:noise}.

The results show that the impact of noise on accuracy is significant when $\epsilon=1$ or $\epsilon=2$. 
However, when $\epsilon \ge 10$, the loss of accuracy is minimal. 
This is because the probability of flipping one bit's sign is 0.33\% when $\epsilon=10$, meaning that accurate information can be maintained, leading to better performance.

Additionally, the loss of accuracy decreases with the increase in hash code length. 
This is expected as adding noise to each bit can be regarded as a Bernoulli trial with a probability of $\frac{1}{2}e^{-\frac{\epsilon}{2}}$. 
Given an expected number $k$ of flipped bits and the length $n$, the probability can be calculated using the formula 
$$
Pr[H(\textbf{h}',\textbf{h})=k]=\binom{n}{k}(\frac{1}{2}e^{-\frac{\epsilon}{2}})^k(1-\frac{1}{2}e^{-\frac{\epsilon}{2}})^{(n-k)}, 
$$
where $H(\cdot)$ calculates the Hamming distance and $\textbf{h}$ is the perturbed code of $\textbf{h}$. For example, when $\epsilon=1$, and taking the case of 16 bits and $k=4$, the corresponding probability is approximately 20\%.
This means that a longer code can retain most of the valid bits, maintaining performance.

In conclusion, we believe that incorporating DP in HashVFL is unnecessary as it would significantly reduce model performance with a small privacy budget and offer limited privacy protection with a large privacy budget.
If DP is deemed necessary, the length of the hash code should be increased accordingly to reduce performance loss.

\begin{table}[t]
\centering
\caption{Performance comparison with different privacy budgets. The cells report the accuracy on the test set.}
\label{table:noise}
\small
\resizebox{\columnwidth}{!}{%
\begin{tabular}{@{}p{2cm}|p{2cm}|l|l|l|l@{}}
\toprule
Dataset & Code Length & $\epsilon=1$ & $\epsilon=2$ & $\epsilon=10$ & $\epsilon=\infty$  \\ \midrule
\multirow{3}{*}{MNIST}   & 4 Bits  & 34.28 & 61.73 & 97.26          & \textbf{97.75} \\ \cmidrule(l){2-6} 
                         & 8 Bits  & 45.65 & 80.34 & 97.95          & \textbf{98.34} \\ \cmidrule(l){2-6} 
                         & 16 Bits & 65.62 & 93.35 & 98.21          & \textbf{98.42} \\ \midrule
\multirow{3}{*}{CIFAR10} & 4 Bits  & 26.89 & 44.76 & 69.11          & \textbf{70.83} \\ \cmidrule(l){2-6} 
                         & 8 Bits  & 33.54 & 57.40 & 72.99          & \textbf{73.65} \\ \cmidrule(l){2-6} 
                         & 16 Bits & 47.08 & 68.73 & 74.23          & \textbf{74.96} \\ \midrule
\multirow{3}{*}{FER}     & 4 Bits  & 27.57 & 39.08 & 50.57          & \textbf{52.73} \\ \cmidrule(l){2-6} 
                         & 8 Bits  & 33.81 & 56.38 & 53.15          & \textbf{54.50} \\ \cmidrule(l){2-6} 
                         & 16 Bits & 40.32 & 51.56 & \textbf{55.06} & 55.00          \\ \midrule
\multirow{3}{*}{CBPD}    & 4 Bits  & 58.56 & 66.06 & 70.38          & \textbf{70.98} \\ \cmidrule(l){2-6} 
                         & 8 Bits  & 61.99 & 67.85 & 75.68          & \textbf{76.62} \\ \cmidrule(l){2-6} 
                         & 16 Bits & 63.96 & 69.44 & 76.64          & \textbf{77.35} \\ \midrule
\multirow{3}{*}{CRITEO}  & 4 Bits  & 64.04 & 70.18 & \textbf{73.82} & 73.70          \\ \cmidrule(l){2-6} 
                         & 8 Bits  & 64.55 & 71.03 & 72.74          & \textbf{73.13} \\ \cmidrule(l){2-6} 
                         & 16 Bits & 66.81 & 72.31 & \textbf{74.03} & 73.32          \\ \midrule
\multirow{3}{*}{IMDb}    & 4 Bits  & 64.03 & 68.51 & \textbf{69.35} & 68.59          \\ \cmidrule(l){2-6} 
                         & 8 Bits  & 66.61 & 69.66 & 71.44          & \textbf{72.10} \\ \cmidrule(l){2-6} 
                         & 16 Bits & 68.26 & 70.21 & 72.42          & \textbf{72.66} \\ \bottomrule
\end{tabular}%
}
\end{table}
\section{Sensitivity Analysis}
\label{sec:sensitivity}
This section delves deeper into the effects of the default setting in previous experiments. 
It addresses the following research questions:
\begin{itemize}
    \item \textbf{Q1:} How does the \textit{number of parties} affect HashVFL?
    \item \textbf{Q2:} How does the \textit{length of hash codes} affect HashVFL?
    \item \textbf{Q3:} How does different \textit{feature ratios} affect HashVFL?
    \item \textbf{Q4:} How does the \textit{number of classes} affect HashVFL?
\end{itemize}

\subsection{Number of Parties (Q1)}\label{sec:parties}
This section explores the impact of the number of parties on performance. 
We split the features of samples in a dataset to simulate multi-party scenarios, resulting in each party having fewer features as the number of parties increases.
Considering that many columns in images of MNIST are black, we decide to exclude it from the datasets used in the experiment.
We conduct experiments on CIFAR10, FER, CBPD, and CRITEO, but not on IMDb as it requires too much memory to run BERTs simultaneously.

\begin{figure}[t]
    \centering
    \includegraphics[width=\columnwidth]{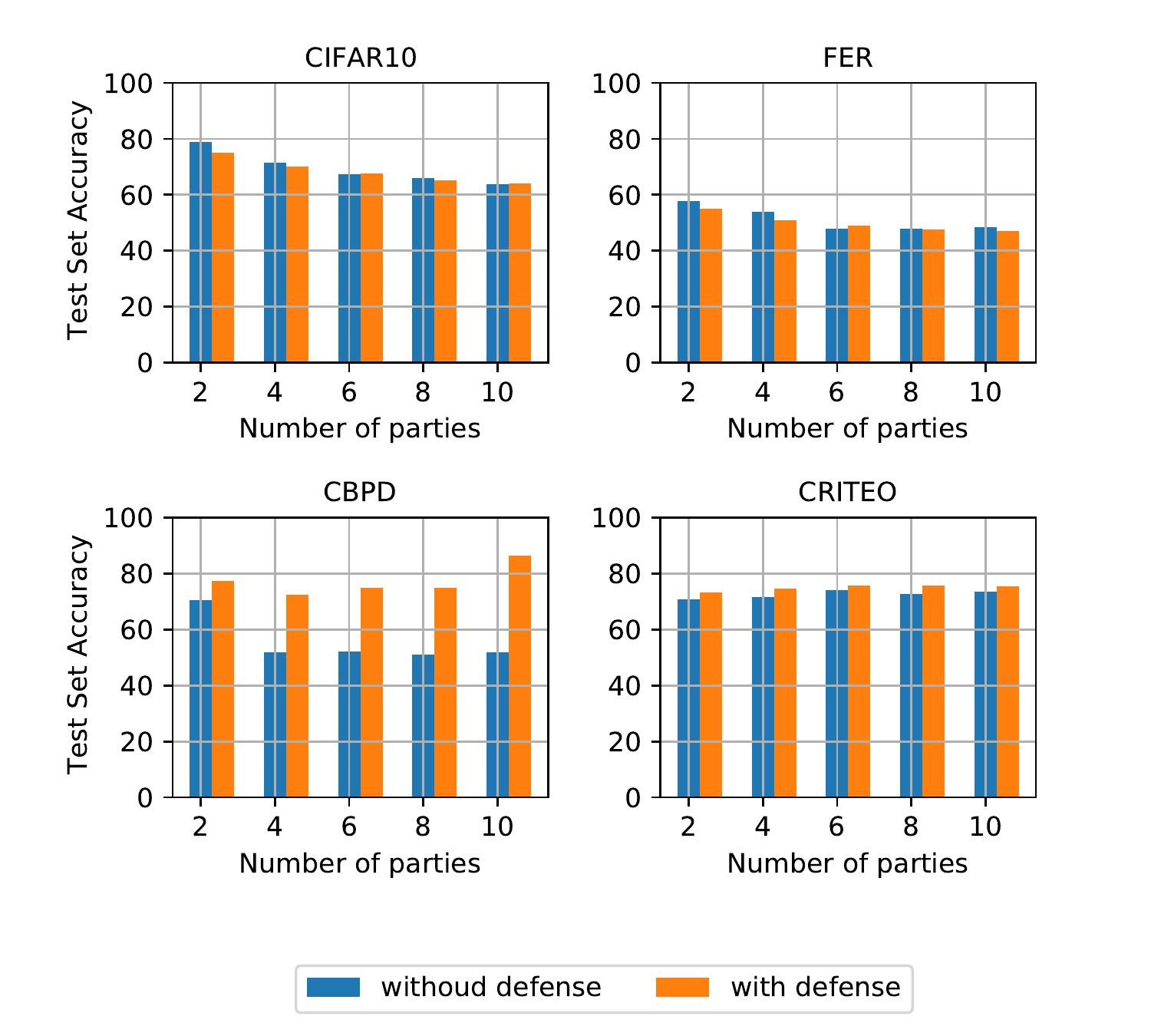}
    \caption{Impact of number of parties on accuracy. X-axis is the number of parties involved, while the Y-axis is the accuracy achieved on the test set.}
    \label{fig:classes}
\end{figure}

Each party holds the same feature ratio, rounded to the nearest whole number. 
For example, with 3 parties, the ratios are 30\%, 30\%, and 40\%. 
To mitigate the loss in accuracy, we set the length of hash codes to 16 bits. 
The results are shown in Fig.~\ref{fig:classes}.

On CIFAR10 and FER, performance decreases with the increasing number of parties, as splitting useful features can destroy their integrity. 
In contrast, on CBPD and CRITEO, performance remains stable as tabular data features are more independent.

Despite the number of parties affecting main task performance, our proposed HashVFL framework maintains performance close to without defense, and even improves performance for tabular datasets as seen in Section~\ref{sec:performance}.

In conclusion, while the number of parties impacts performance, HashVFL can maintain close to the performance without defense.

\subsection{Length of Hash Codes (Q2)}
In this section, we assess the impact of varying the length of hash codes on performance. 
By doubling the length of hash codes from 4 bits to 128 bits, we observe improved performance initially, which then converges. 
TABLE~\ref{table:length} summarizes the results. 
We speculate that the improvement in the previous stage is because the increased bits can compensate for the information loss caused by hashing, and when the information that the model can extract is saturated, more bits can only cause redundancy.

It is recommended to determine the appropriate hash code length according to the required security level, where longer hash codes result in improved performance and shorter hash codes offer stricter data protection.

\begin{table}[t]
\centering
\caption{Performance comparison with different lengths. The cell reports the accuracy on the test set.}
\label{table:length}
\resizebox{\columnwidth}{!}{%
\begin{tabular}{l|c|c|c|c|c|c}
\toprule
Dataset & 4 Bits    & 8 Bits    & 16 Bits   & 32 Bits   & 64 Bits   & 128 Bits    \\ \midrule
MNIST   & 97.75     & 98.34     & 98.42     & 98.59     & 98.42     & 98.57 \\ \midrule
CIFAR10 & 70.83     & 73.65     & 74.96     & 76.14     & 75.34     & 75.03 \\ \midrule
FER     & 52.73     & 54.50     & 55.00     & 55.66     & 55.34     & 55.61 \\ \midrule
CBPD    & 70.98     & 76.62     & 77.35     & 77.95     & 81.64     & 85.00 \\ \midrule
CRITEO  & 73.70     & 73.13     & 73.32     & 74.08     & 74.48     & 74.67 \\ \midrule
IMDb    & 68.59     & 72.10     & 72.66     & 73.10     & 73.46     & 73.59 \\ \bottomrule
\end{tabular}%
}
\end{table}

\subsection{Feature Ratio (Q3)}\label{sec:feature}
This section evaluates the effect of the feature ratio, which is defined as the proportion of features owned by a single party in the whole set of features. 
Experiments were conducted in a two-party scenario, where one party's feature ratio varied from 10\% to 50\%. 
The symmetric scenario of varying feature ratio from 60\% to 90\% was omitted as it was expected to be the same. 

\begin{figure}[t]
    \centering
    \includegraphics[width=\columnwidth]{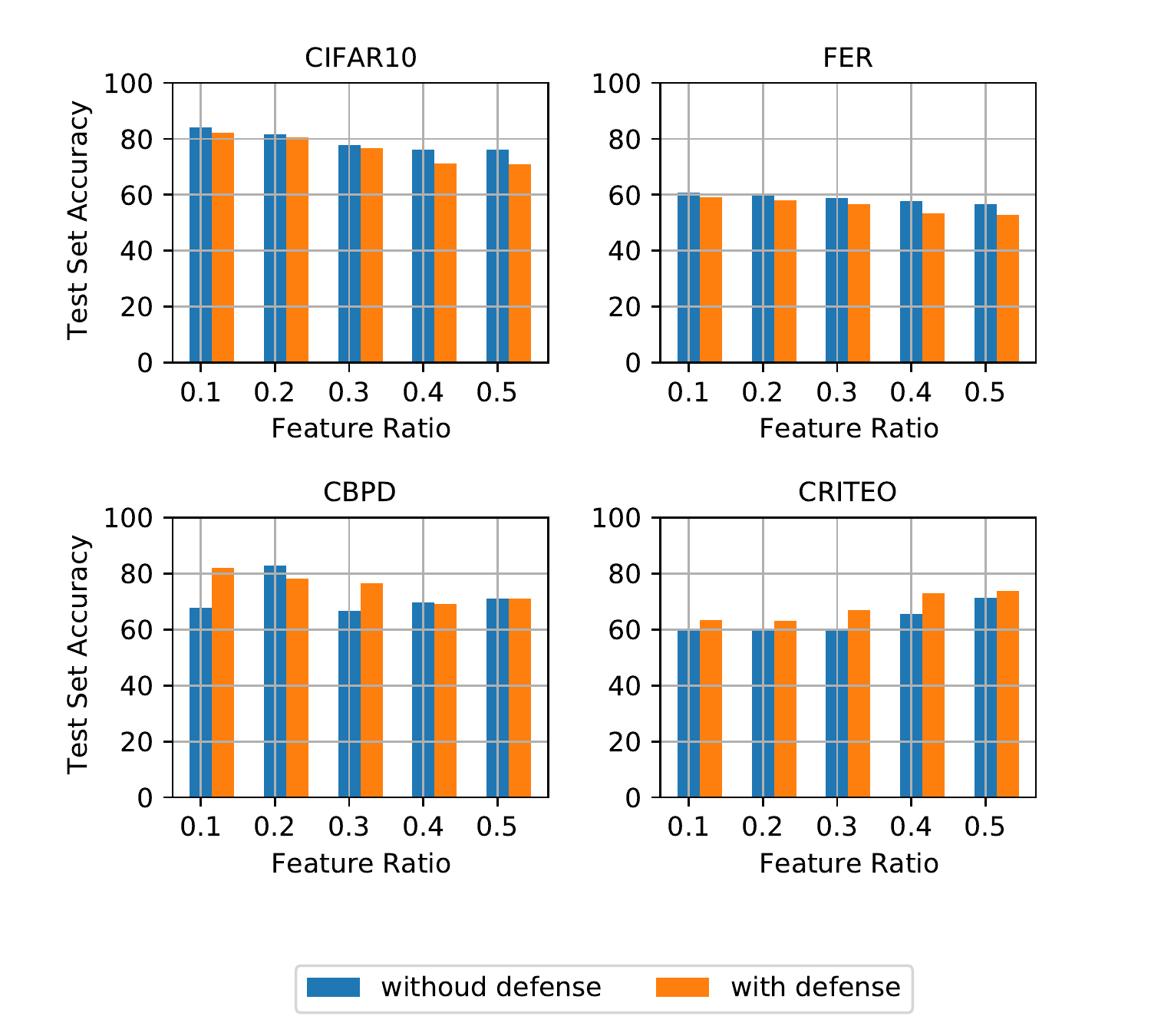}
    \caption{Impact of feature ratio on accuracy. X-axis is the ratio of features held by one party, while the Y-axis is the accuracy achieved on the test set.}
    \label{fig:ratio}
\end{figure}

The results, shown in Fig.~\ref{fig:ratio}, indicated that performance decreases with increasing feature ratio on CIFAR10 and FER. 
This is because as the feature ratio increases, the other party's image completeness decreases, and important features become concentrated in the middle region, making inference difficult. 
However, on CBPD and CRITEO, where features are more independent, the effect of the feature ratio was less pronounced with some fluctuations. 

The results showed that HashVFL kept the accuracy loss within an acceptable range on CIFAR10 and CRITEO and even improved performance on CBPD and CRITEO, as analyzed in Section~\ref{sec:performance}. 
In conclusion, complete and valid features are crucial for reducing accuracy loss in HashVFL.

\subsection{Number of Classes (Q4)}
The results of our experiments on CIFAR10 and CIFAR100, which share the same data but have different number of classes, showed that the number of classes does make the task more challenging. 
As seen in TABLE~\ref{table:classes}, the highest accuracy achieved by our models on the CIFAR100 test set was 33.77\% when the length was 64 bits with defense, while the accuracy on CIFAR10 was 75.34\% with the same length.

When the length was short, the accuracy loss was higher on CIFAR100 compared to CIFAR10. 
For instance, when the length was 8 bits, the accuracy loss on CIFAR100 was 10\%, while it was 6\% on CIFAR10 (as shown in TABLE~\ref{table:performance}). 
However, when the length was increased to 32 or 64 bits, the loss of accuracy was less than 1\%.

In conclusion, a larger number of classes increases the complexity of the task and causes a significant accuracy loss when the length of the hash codes is short. 
But, when there are enough bits, our method, HashVFL, can maintain acceptable performance.

\begin{table}[t]
\centering
\caption{Performance comparison with different numbers of classes. We plot the results of CIFAR10 with defense for reference.}
\label{table:classes}
\small
\resizebox{\columnwidth}{!}{%
\begin{tabular}{@{}l|l|c|c|c|c|c@{}}
\toprule
Dataset  &  & 8 Bits      & 16 Bits      & 32 Bits       & 64 Bits       & 128 Bits    \\ \midrule
\multirow{2}{*}{CIFAR100} & without defense & 33.99     & 34.09   & 34.75  & 33.39  & 34.41\\\cmidrule(l){2-7}
                          & with defense    & 24.23     & 29.51   & 33.31  & 33.77  & 31.96 \\\midrule
CIFAR10                   & with defense    & 73.65     & 74.96   & 76.14  & 75.34  & 75.03 \\ \bottomrule
\end{tabular}%
}
\end{table}
\section{Ablation Study}
\label{sec:ablation}
Three challenges were introduced in Section~\ref{sec:introduction}. While previous works like \cite{Wu2019hashing, Cao2017HashNetDL, su2018greedyhash, Pham2022BinarizingSL} have addressed the challenge of learnability in deep hashing using different approaches, we find that the combination of the $Sign$ function and the STE is the most efficient and completely irreversible during training.
Other functions like $tanh$ and $sigmoid$ preserve the values during training, leaving the risk of information leakage.

However, while this technique has been proven effective in retrieval systems \cite{su2018greedyhash} and split learning \cite{Pham2022BinarizingSL}, there are two additional challenges, namely bit balance and consistency, that need to be addressed in the context of VFL.
In this section, our primary objective is to answer two important questions: the necessity of the incorporated modules in addressing the challenges we encountered, and the effectiveness of these modules compared to Greedy Hash \cite{su2018greedyhash} and B-SL \cite{Pham2022BinarizingSL}. We summarize these questions as follows:
\begin{itemize}
    \item \textbf{AS1:} What is the role of \textit{bit balance} in HashVFL?
    \item \textbf{AS2:} What benefit does \textit{consistency} bring to HashVFL?
\end{itemize}

\subsection{Bit Balance (AS1)}\label{sec:bn}
This section provides experimental evidence for the importance of the BN layer in HashVFL to address the challenge of bit balance.

\textbf{Baseline:} We compare our HashVFL approach with Greedy Hash. Greedy Hash also uses the $Sign$ function and STE for gradient estimation, but its design focuses on improving retrieval performance and reducing the impact of hash collisions. 
Consequently, it does not consider the need to maximize the leverage of each bit, as its hash code's length can exceed 128 bits. 
Additionally, Greedy Hash introduces a penalty term based on the Euclidean distance between the embeddings and their corresponding binary codes.

\textbf{Experimental Setup:} By default, we conduct experiments in a two-party scenario, where each party holds half of the features. We vary the length of the hash code from 4 bits to 16 bits for comparison.

The results, summarized in TABLE~\ref{table:bn}, demonstrate that our method with a BN layer outperforms Greedy Hash on all datasets, while performing similarly without it. 
On CBPD, CRITEO, and IMDb datasets, our method without a BN layer exhibits a significant loss of performance, but adding a BN layer mitigates this issue.

We speculate that the BN layer reduces the impact of large rotations caused by the cosine similarity loss during optimization. 
For instance, in a binary classification task with a 4-bit hash code length, there are $2^4\cdot2=32$ rotation angles involved in the optimization process. 
Compared to the changes in gradients resulting from the classification loss, the rotation caused by the similarity loss may be too large, causing the hash code of a sample to flip at every optimization step. 
Consequently, it becomes difficult for the top model to learn a stable function for accurate prediction. 
However, the addition of a BN layer can alleviate this issue by reducing the impact of rotations. 
The BN layer evenly divides the distribution of each bit for every batch, allowing for a larger range for each bit to vary without flipping its sign.

In conclusion, simply applying the approach used in Greedy Hash cannot address the performance degradation in VFL when using limited-length hash codes. 
The BN layer is essential in HashVFL as it evenly divides the distribution of each bit, thereby maximizing the leverage of each bit.

\begin{table}[t]
\centering
\caption{Ablation study of batch normalization's impact. `Greedy Hash' denotes the baseline for reference. `Ours' refers to our design. The cell reports the accuracy on the test set.}
\label{table:bn}
\small
\resizebox{\columnwidth}{!}{%
\begin{tabular}{p{2cm}|p{2cm}|l|l|l|l}
\toprule
Dataset                  & Method                 &     & 4 Bits & 8 Bits & 16 Bits \\ \midrule
\multirow{3}{*}{MNIST}   & Greedy Hash            &     & 96.24  & 97.17  & 97.82   \\ \cmidrule(l){2-6}
                         & \multirow{2}{*}{Ours} & without BN & 96.75  & 97.24  & 97.02   \\ \cmidrule(l){3-6}
                         &                       & with BN    & 97.75  & 98.34  & 98.42    \\ \midrule
\multirow{3}{*}{CIFAR10} & Greedy Hash            &     & 55.53  & 63.57  & 61.52   \\  \cmidrule(l){2-6}
                         & \multirow{2}{*}{Ours} & without BN & 60.44  & 63.24  & 60.26   \\ \cmidrule(l){3-6}
                         &                       & with BN    & 70.83  & 73.65  & 74.96    \\ \midrule
\multirow{3}{*}{FER}     & Greedy Hash            &     & 40.42  & 42.24  & 44.65   \\  \cmidrule(l){2-6}
                         & \multirow{2}{*}{Ours} & without BN & 37.95  & 45.74  & 48.68   \\ \cmidrule(l){3-6} 
                         &                       & with BN   & 52.73  & 54.50  & 55.00    \\ \midrule
\multirow{3}{*}{CBPD}    & Greedy Hash            &     & 61.57  & 63.13  & 62.42   \\  \cmidrule(l){2-6}
                         & \multirow{2}{*}{Ours} & without BN & 48.74  & 49.89  & 48.56   \\  \cmidrule(l){3-6}
                         &                       & with BN   & 70.98  & 76.62  & 77.35    \\ \midrule
\multirow{3}{*}{CRITEO}  & Greedy Hash            &     & 63.21  & 66.94  & 68.63   \\  \cmidrule(l){2-6}
                         & \multirow{2}{*}{Ours} & without BN & 49.76  & 49.93  & 49.94   \\ \cmidrule(l){3-6}
                         &                       & with BN   & 73.70  & 73.13  & 73.32    \\ \midrule
\multirow{3}{*}{IMDb}    & Greedy Hash            &     & 70.71  & 70.63  & 70.55   \\ \cmidrule(l){2-6}
                         & \multirow{2}{*}{Ours} & without BN & 50.59  & 50.34  & 50.26   \\ \cmidrule(l){3-6}
                         &                       & with BN   & 68.59  & 72.10  & 72.66    \\ \bottomrule
\end{tabular}%
}
\end{table}

\subsection{Consistency (AS2)}
When comparing VFL to split learning, it is important to consider the computational costs associated with an increasing number of parties.
In Section~\ref{sec:introduction}, we introduced the challenge of consistency in VFL and proposed the use of predefined target binary codes to reduce the complexity of calculating distances between parties from $O(N^2)$ to $O(N)$. 
This approach saves computational resources and accelerates training.

In addition to the computational benefits, we are also interested in exploring the additional advantages of consistency in HashVFL. We speculate that consistency can simplify the task by reducing the number of combinations of multi-party hash codes.

\textbf{Baseline:} To establish a baseline comparison, we choose an extension of B-SL as our baseline, which does not utilize predefined binary codes. Additionally, we introduce an Euclidean distance penalty in B-SL to ensure model convergence without consistency guarantees.

\textbf{Experimental Setup:} We conduct the experiments in a two-party scenario, with each party holding half of the features. Furthermore, we still vary the length of the hash code from 4 bits to 16 bits for comparison.

\begin{figure}
    \centering
    \includegraphics[width=\columnwidth]{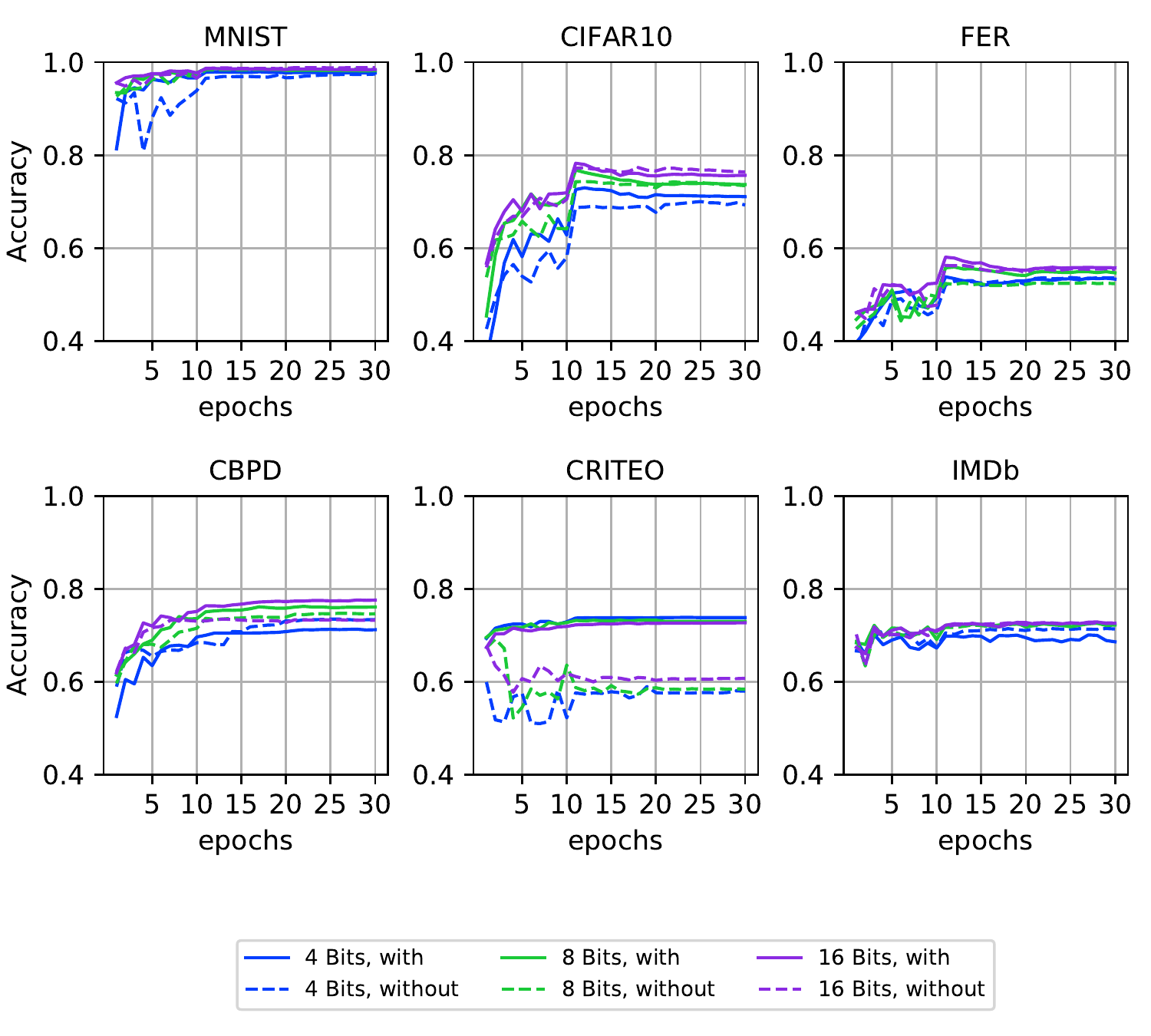}
    \caption{Comparison of performance with and without addressing `consistency'. 
    The X-axis represents the training epoch, and the Y-axis shows the accuracy. 
    The legends `with' and `without' indicate the presence or absence of cosine similarity.}
    \label{fig:consistency}
\end{figure}

The results, as depicted in Fig.~\ref{fig:consistency}, demonstrate that the curves with consistency requirements outperformed those without on all datasets, with a gap of nearly 15\% on CRITEO and around 1\% on other datasets, except for CBPD and IMDb with 4-bit codes.
Moreover, the inclusion of the cosine similarity loss also accelerated training, resulting in a faster improvement in accuracy during the initial 10 epochs. 
However, after 10 epochs, the accuracy on the training set continued to increase while the accuracy on the test set decreased, potentially indicating overfitting. 
By the 30-th epoch, the accuracy on the test set decreased for training with cosine similarity loss, while it improved for training without it.

In conclusion, addressing consistency in HashVFL not only provides additional benefits such as detecting abnormal inputs and reducing computational costs, but also improves and accelerates training. 
These findings underscore the significance of consistency in HashVFL.

\section{Related Work}
\label{sec:related}
In this section, we supplement the related work and compare the work we refer to with our method.

\subsection{Learning to Hash}
Nearest neighbor search is a problem that seeks to find the samples in a database with the smallest distance to the query. Hashing is a commonly used solution due to its computational and storage efficiency. 
With the advancement of deep neural networks (DNNs), deep hashing has emerged as a more effective solution than traditional methods. 
Deep supervised hashing \cite{Wu2019hashing,su2018greedyhash,Cao2017HashNetDL,Wang2018hashing,Liu2016DeepSH,Li2016FeatureLB,Jiang2018AsymmetricDS,hoe2021one,fan2020dpn} is a subfield of deep hashing, where the goal is to solve binary optimization and address vanishing gradients in DNNs.

Different approaches have been proposed in the literature for deep supervised hashing. 
A line of work focuses on binarizing the activations in DNNs. 
For example, Cao et al. \cite{Cao2017HashNetDL} uses a combination of $Tanh$ and $Sigmoid$ functions, while Li et al. \cite{Li2016FeatureLB} designed a penalty function to generate binary features. 
In \cite{su2018greedyhash}, Su et al. proposed the $Sign$ function to directly binarize the features.

Another line of work focuses on a different approach to learning hashing. 
For example, Fan et al. \cite{fan2020dpn} used a random assignment scheme to generate target vectors with maximal inter-class distance.
Then, they optimized the distance between the embeddings and the vectors.
Yuan et al. \cite{Yuan2020CentralSQ}, however, used the Hadamard matrix as target centers. 
Hoe et al. \cite{hoe2021one} integrated category information into one loss by revealing the connection between cosine similarity and Hamming distance. 

\subsection{Attacks in Vertical Federated Learning}
Recently, several studies have explored the security of VFL. 
These studies mainly focus on two aspects: data privacy and security.

Regarding privacy, Luo et al. \cite{luo2020feature} proposed a DNN-based method for reconstructing data in VFL. 
Weng et al. \cite{weng2021privacy} also studied the privacy risks of VFL using machine learning methods such as logistic regression and XGBoost. 
Qiu et al. \cite{qiu2022relation} investigated the privacy risks of graph data in VFL, and Fu et al. \cite{fu2022label} looked into the leakage of labels.

For security, Liu et al. \cite{Liu2021BatchLI} found that the party that owns the label can easily carry out a backdoor attack. 
They also explored the possibility of a backdoor attack when the adversary has no access to labels and found that replacing gradients can be effective.

\subsection{Defenses in Vertical Federated Learning}
VFL is a relatively new field and there have been limited studies on defenses against attacks in VFL. 
Two lines of research have been proposed to address different types of attacks.

In \cite{sun2021defend} and \cite{Vepakomma2020NoPeekIL}, Sun et al. and Vepakomma et al. proposed schemes to reduce data reconstruction attacks by incorporating the correlation distance between extracted embeddings and raw inputs into the penalty function. 
Sun et al. proposed a method to defend against label inference attacks by integrating DP into the forward process in \cite{Sun2022LabelLA}.
Defenses against label inference attacks were also discussed in \cite{fu2022label} by using gradient compression \cite{Lin2018DeepGC}.
Pham et al. proposed a defense against feature reconstruction attacks by integrating binary neural networks (BNNs) \cite{Qin2020BinaryNN} into the first few layers in \cite{Pham2022BinarizingSL}.

In \cite{Liu2021RVFRRV}, Liu et al. used feature reconstruction to defend against backdoor attacks, which applied an attack for good.

\subsection{Remark}
Our proposed HashVFL aims to defend against feature reconstruction attacks through the use of hashing. 
Unlike prior defenses such as \cite{sun2021defend,Vepakomma2020NoPeekIL}, hashing can eliminate the connection between the binarized embedding and the input even when the adversary has complete knowledge of the model and the hash code. 

Comparing to the defense proposed by Pham et al. \cite{Pham2022BinarizingSL}, who only binarizes the first few layers of the feature maps, HashVFL is capable of handling more complex scenarios, such as different types of data, and can easily be integrated into existing frameworks.

Research in learning to hash has provided valuable insights for integrating hashing into our design. 
The GreedyHash method proposed by Su et al. \cite{su2018greedyhash} offers scalability suitable for VFL model. 
However, it does not fully address the challenges of balancing bits and consistency. 

Methods such as \cite{Cao2017HashNetDL,Li2016FeatureLB} are effective in learning to hash, but they still entail the risk of leakage ($Tanh$ and $Sigmoid$ leave the risk of reversibility). 
Hence, we do not adopt these methods in HashVFL's design. 

\section{Discussion}\label{sec:discussion}
\subsection{Adaptive Attack}
Our evaluation demonstrates that HashVFL effectively mitigates the privacy issues stemming from reconstruction attacks.
However, as identified in the analysis of label leakage, it is possible to infer certain attributes with a specific classifier.
Considering the remarkable performance of generative models like ChatGPT \cite{OpenAI2023GPT4TR} and Stable Diffusion models \cite{rombach2021highresolution}, we speculate that a feasible adaptive attack to bypass HashVFL could follow the following steps:

\begin{enumerate}
    \item Extracting relevant attributes through PLA using a set of classifiers.
    \item Utilizing the extracted attributes to construct a prompt.
    \item Employing a suitable generative model to reconstruct the target sample using the earlier prompt.
\end{enumerate}

The intuition behind this adaptive attack is to maximize the extraction of common features of a particular class of samples that are retained in HashVFL.

In the future, it is crucial to further decouple data utilization and data distribution based on these findings. 
This can help strengthen the defense mechanism against adaptive attacks and ensure that data privacy is upheld effectively.

\subsection{Bias Between Parties}
Section~\ref{sec:feature} reveals that when one party possesses a majority of features, the performance improves compared to that both parties hold equal portions. 
Our analysis attributes this improvement to feature completeness in the former scenario, enabling the top model to effectively use the information.
However, this also raises a concern about the party with more features having greater influence on VFL's final predictions. 
The use of HashVFL may exacerbate this bias by discarding much of the information for all parties.

The presence of the bias in the dominant party raises the concern of malicious manipulation of the final prediction during inference in HashVFL. 
Addressing this bias and promoting equal feature importance in the prediction is a key challenge for its practical implementation. 

Our HashVFL design has been proven effective in detecting malicious code through consistency checks. 
However, how to mitigate such bias in training remains an open area for future research with promising potential.

\section{Conclusion}\label{sec:conclusion}
This work introduces HashVFL, a new VFL framework that leverages hashing to defend against data reconstruction attacks. 
As far as we know, this is the first VFL framework that incorporates hashing. 
We address three challenges in integrating hashing into VFL and provide effective solutions. 
Our evaluation results show that HashVFL retains the performance of the main task while effectively protecting against data reconstruction attacks. 
Additionally, we show experimentally that HashVFL can reduce the degree of label leakage, mitigate the adversarial attack, and detect abnormal inputs. 
We anticipate that this work will spark further investigations into the practical applications of HashVFL.

\section*{Acknowledgments}
This work was partly supported by the National Key Research and Development Program of China under No. 2022YFB3102100 and NSFC under No. 62102360.

\bibliographystyle{IEEEtran}
\bibliography{ref}

\begin{IEEEbiography}[{\includegraphics[width=1in,height=1.25in,clip,keepaspectratio]{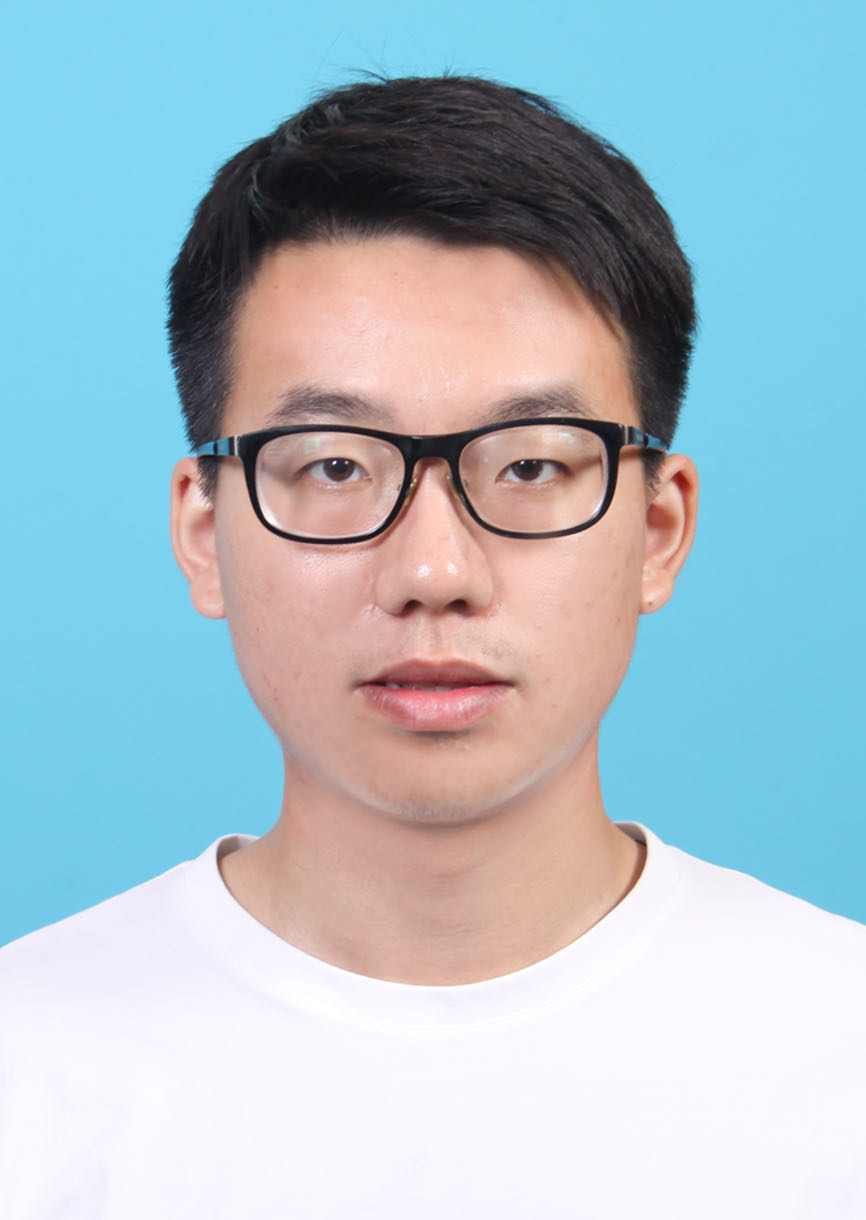}}]{Pengyu Qiu}
is currently a Ph.D. student in the College of Computer Science and Technology at Zhejiang University. He received his Bachelor's degree from Zhejiang University. His current research interests include Federated Learning, AI security.
\end{IEEEbiography}

\begin{IEEEbiography}[{\includegraphics[width=1in,height=1.25in,clip,keepaspectratio]{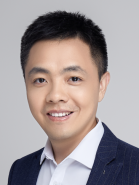}}]{Xuhong Zhang}
is a ZJU 100-Young Professor with the School of Software Technology at Zhejiang University. He received his Ph.D. in Computer Engineering from University of Central Florida in 2017 . His research interests include distributed big data and AI systems, big data mining and analysis, data-driven security, AI and Security. He has authored over 20 publications in premier journals and conferences such as TDSC, TPDC, IEEE S\&P, USENIX Security, ACM CCS, NDSS, VLDB, etc.
\end{IEEEbiography}

\begin{IEEEbiography}[{\includegraphics[width=1in,height=1.25in,clip,keepaspectratio]{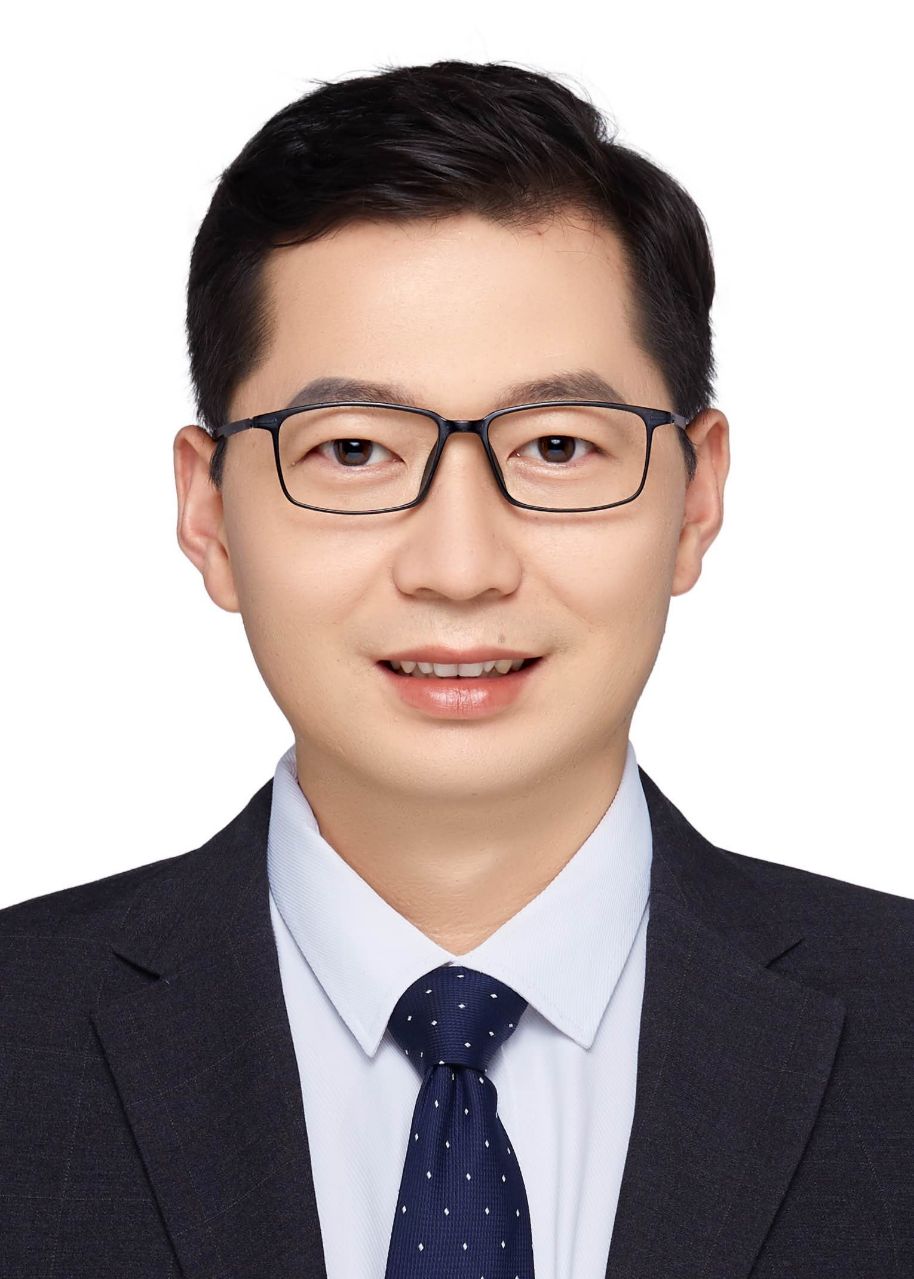}}]{Shouling Ji}
is a Qiushi Distinguished Professor in the College of Computer Science and Technology at Zhejiang University. He received a Ph.D. degree in Electrical and Computer Engineering from Georgia Institute of Technology and a Ph.D. degree in Computer Science from Georgia State University. His current research interests include Data-driven Security and Privacy, AI Security and Software and System Security. He is a member of ACM and IEEE, and a senior member of CCF. He was a Research Intern at the IBM T. J. Watson Research Center. Shouling is the recipient of the 2012 Chinese Government Award for Outstanding Self-Financed Students Abroad and 10 Best/Outstanding Paper Awards, including ACM CCS 2021.
\end{IEEEbiography}

\begin{IEEEbiography}[{\includegraphics[width=1in,height=1.25in,clip,keepaspectratio]{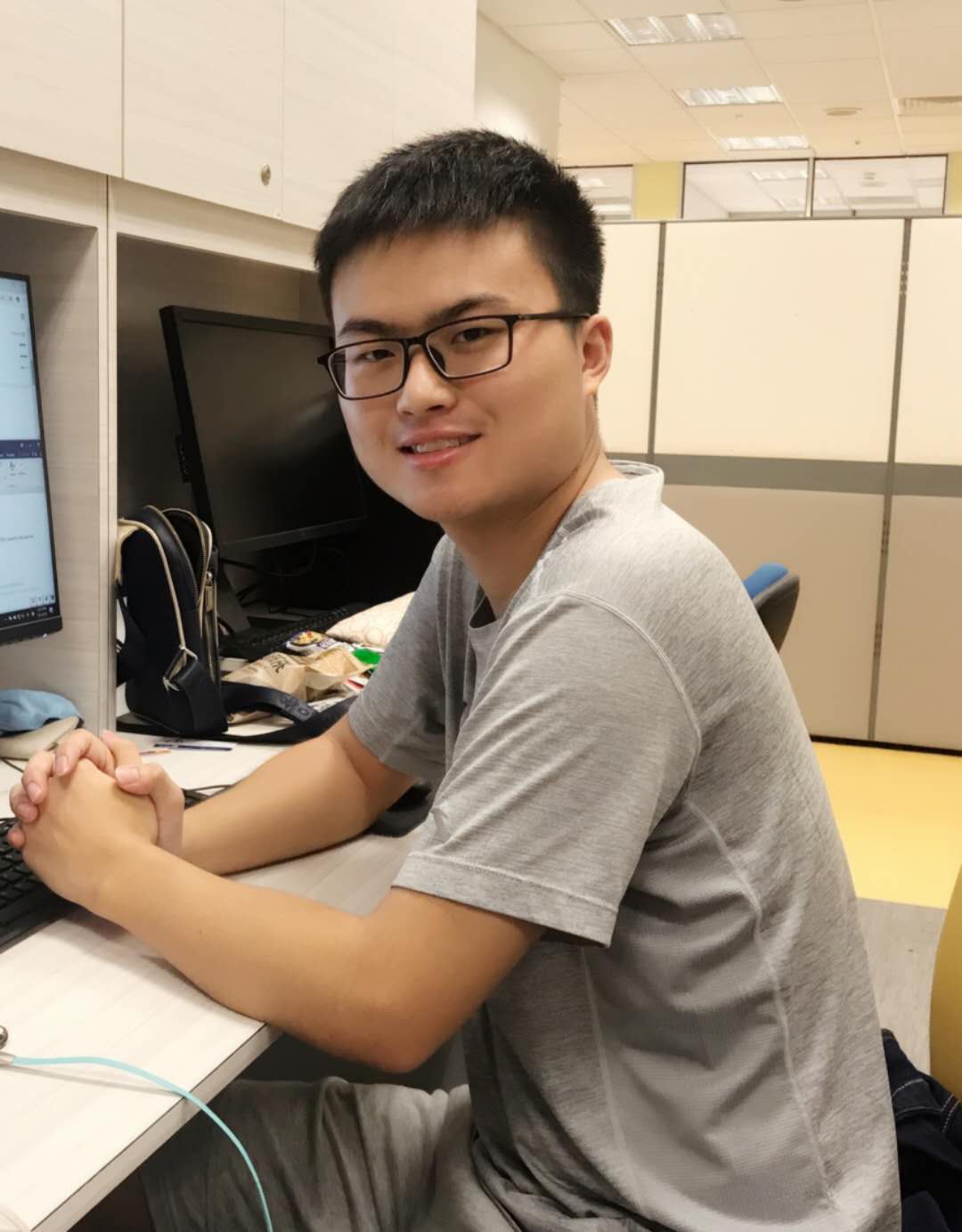}}]{Chong Fu} 
is currently a Ph.D. student in the College of Computer Science and Technology at Zhejiang University. He received his Bachelor's degree from Jilin University. His current research interests include federated learning and adversarial machine learning.
\end{IEEEbiography}

\begin{IEEEbiography}[{\includegraphics[width=1in,height=1.25in,clip,keepaspectratio]{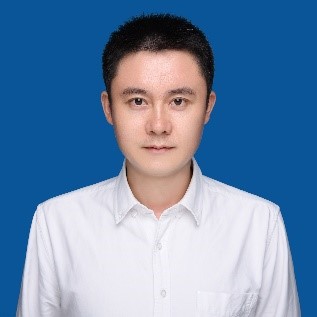}}]{Xing Yang} 
is a researcher at the State Key Laboratory of Pulsed Power Laser Technology, National University of Defense Technology. He received his BS, MS and Ph.D. degrees from Hefei Electronic Engineering Institute in 2006, 2009, and 2012 respectively. Currently, his research interests mainly focus on optoelectronic engineering, artificial intelligence, and cyberspace security.
\end{IEEEbiography}

\begin{IEEEbiography}[{\includegraphics[width=1in,height=1.25in,clip,keepaspectratio]{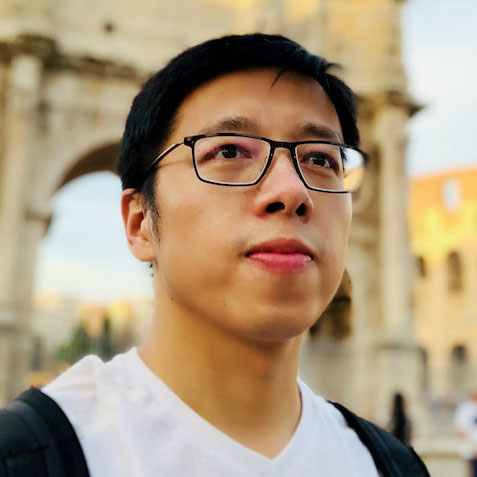}}]{Ting Wang} 
is an assistant professor in the College of Information Sciences and Technology at Penn State. He received his Ph.D. degree from Georgia Tech. He conducts research at the intersection of data science and privacy \& security. His ongoing work focuses on making machine learning systems more practically usable through improving their Security Assurance, Privacy Preservation and Decision-Making Transparency.
\end{IEEEbiography}

\end{document}